\documentclass[a4paper]{article}
\pdfoutput=1 

\usepackage{jheppub} 

\usepackage{graphicx,epsfig,wrapfig,amssymb,color,amsmath,bm,breqn}

\usepackage[T1]{fontenc} 
\usepackage{multirow}
\usepackage{amsmath}
\usepackage{pifont}
\usepackage{float}

\usepackage{soul}
\usepackage[utf8]{inputenc} 

\usepackage{lineno} 

 \usepackage{hyperref} 
\hypersetup{
    colorlinks,
    citecolor=black,
    filecolor=black,
    linkcolor=black,
    urlcolor=black
}

\newcommand{\bmax}{b_{\rm max}}
\newcommand\bstarsc{b_*}
\newcommand\mubstar{\mu_{\bstarsc}}

\newcommand{\bea}{\begin{eqnarray}}
\newcommand{\eea}{\end{eqnarray}}

\newcommand{\nn}{\nonumber\\}
\newcommand{\be}{\begin{equation}}
\newcommand{\ee}{\end{equation}}
\newcommand{\ba}{\begin{eqnarray}}
\newcommand{\ea}{\end{eqnarray}}
\newcommand{\la}{\langle}
\newcommand{\ra}{\rangle}
\newcommand{\di}{ {\rm d} }

\definecolor{darkgreen}{rgb}{0,0.65,0}
\definecolor{orange}{rgb}{1,0.4,0.1}

\newcommand{\remove}[1]{{   }}

\def\qT{\bm q_T}

\def\kTh{\bm k_{Th}} 
\def\vkTpi{\vec{\bm k}_{T\pi}} 
\def\vkTN{\vec{\bm k}_{Tp}} 
 
\def\kTpi{{\bm k}_{T \pi}} 
\def\kTN{{\bm k}_{Tp}} 

\def\hhat{\hat{\bm h}}

\def\avkTh{\la k_{Th}^2 \ra}
\def\avkTpi{\la k_{T \pi}^2 \ra}
\def\avkTN{\la k_{Tp}^2 \ra}
\def\avqT{\la q_T^2 \ra}

\newcommand*{\FigPath}{./Figs}%
\newcommand*{\BibPath}{.}%

\usepackage[normalem]{ulem}

\title{The Drell-Yan process with pions \\ and polarized nucleons}

\author[a]{S.~Bastami}
\author[b]{L.~Gamberg}
\author[c,d]{B.~Parsamyan}
\author[e,f]{B.~Pasquini}
\author[b,g]{A.~Prokudin}
\author[a]{P.~Schweitzer}

\affiliation[a]{Department of Physics, University of Connecticut,
  Storrs, CT 06269, U.S.A.}
\affiliation[b]{Division of Science, Penn State Berks, Reading,
	PA 19610, USA}
\affiliation[c]{Dipartimento di Fisica, 
  Universit\'a degli Studi di Torino, Italy}
\affiliation[d]{Istituto Nazionale di Fisica Nucleare, 
  Sezione di Torino, Italy}
\affiliation[e]{Dipartimento di Fisica, 
  Universit\'a degli Studi di Pavia, Italy}
\affiliation[f]{Istituto Nazionale di Fisica Nucleare, 
  Sezione di Pavia, Italy}
\affiliation[g]{Thomas Jefferson National Accelerator Facility,
	Newport News, VA 23606, U.S.A.}

\emailAdd{saman.bastami@uconn.edu}
\emailAdd{lpg10@psu.edu}
\emailAdd{bakur@cern.ch}
\emailAdd{barbara.pasquini@unipv.it}
\emailAdd{prokudin@jlab.org}
\emailAdd{peter.schweitzer@uconn.edu}

\abstract{The Drell-Yan process provides important information
on the internal structure of hadrons including transverse
momentum dependent parton distribution functions (TMDs). 
In this work we present calculations for all leading twist structure functions describing the pion induced Drell-Yan process.
The non-perturbative input for the TMDs is taken from the 
light-front  constituent quark model, the 
spectator model, and 
available parametrizations of TMDs extracted from the experimental data. TMD evolution is implemented at Next-to-Leading Logarithmic 
precision for the first time for all asymmetries.
Our results are compatible with the first experimental information, 
help to interpret the data from ongoing experiments, and will
allow one to quantitatively assess the models in future
when more precise data will become available.
}

\notoc 

\begin{document} 

\maketitle

\section{Introduction\label{sec:intro}}

The Drell-Yan (DY) process with pions and nucleons provides
important information on the structure of pion and nucleon. 
The DY differential cross section in the region of low transverse
momentum, $q_T$, of the produced lepton anti-lepton pair is 
subject to the transverse momentum dependent factorization
\cite{Collins:1984kg}.
The corresponding transverse momentum dependent parton distribution functions
(TMDs) \cite{Collins:2011zzd} in the description of DY at low $q_T$  provide
essential information on correlations between transverse parton momenta and
parton or nucleon spin, and describe the three-dimensional structure of
hadrons. Early theoretical studies of TMDs 
in hadron production in proton-proton processes
\cite{Sivers:1989cc,Anselmino:1994tv,Mulders:1995dh} were followed 
by systematic investigations in semi-inclusive deep-inelastic 
scattering (SIDIS) \cite{Cahn:1978se,Kotzinian:1994dv,
Kotzinian:1995cz,Bacchetta:2006tn} and DY
\cite{Tangerman:1994eh,Boer:1997nt,Arnold:2008kf}
(also fragmentation functions \cite{Metz:2016swz}
enter the description of SIDIS). 
The basis for these descriptions are QCD factorization theorems
\cite{Collins:1981uk,Collins:1984kg,Qiu:1991pp,Ji:2004wu,Ji:2006br,Ji:2006vf,
Collins:2011zzd,Aybat:2011zv,Ma:2013aca,Collins:2014jpa,Collins:2016hqq}.

One of the challenges when interpreting pion-induced DY data is the limited 
knowledge of the pion structure. At twist-2 the process is described by
the proton TMDs: unpolarized distribution $f_{1,p}^a$, transversity
distribution $h_{1,p}^a$, Sivers distribution function $f_{1T,p}^{\perp a}$, 
Boer-Mulders distribution $h_{1,p}^{\perp a}$, Kotzinian-Mulders distribution
$h_{1L,p}^{\perp a}$, and ``pretzelosity'' distribution $h_{1T,p}^{\perp a}$,
and pion TMDs:  unpolarized distribution  $f_{1,\pi}^a$, 
Boer-Mulders distribution $h_{1,\pi}^{\perp a}$.

On the proton side, for $f^a_{1,p}$ both collinear and TMD distributions
are well-known \cite{Gluck:1991ng,Gluck:1994uf,Gluck:1998xa,Martin:2009iq,
Harland-Lang:2014zoa,Dulat:2015mca,Landry:2002ix,Anselmino:2013lza,
Signori:2013mda,Bacchetta:2019sam,Scimemi:2019cmh}.  
Based on global QCD analyses of data, parametrizations are available also for
$f_{1T,p}^{\perp a}$, $h_{1,p}^a$, $h_{1,p}^{\perp a}$, $h_{1T,p}^{\perp a}$ \cite{Anselmino:2011gs,Anselmino:2013vqa,Barone:2009hw,Lefky:2014eia,Cammarota:2020qcw}. 
Only $h_{1L,p}^{\perp a}$ has not yet been extracted, 
though it can be described based on $h_{1,p}^a$ in the 
so-called Wandzura-Wilczek- (WW-)type approximation 
which is compatible with available data~\cite{Bastami:2018xqd}.
On the pion side the situation is different. While extractions of $f_{1,\pi}^a$ 
exist \cite{Gluck:1991ey,Sutton:1991ay,Gluck:1999xe,Aicher:2010cb,Barry:2018ort,Novikov:2020snp}, 
no results on $h_{1,\pi}^{\perp a}$ are available. This constitutes 
a ``bottleneck'' if one would like to describe the pion-induced DY data, 
e.g.\ COMPASS results \cite{Aghasyan:2017jop}, 
based solely on phenomenological extractions since 
$h_{1,\pi}^{\perp a}$ is relevant for the majority of observables 
in the pion-induced polarized DY process at leading twist.
In this situation we will resort to model studies of the pion Boer-Mulders function 
$h_{1,\pi}^{\perp a}$.

An important goal of theoretical studies in models is to describe 
hadron structure at a low initial scale $\mu_0< 1\,{\rm GeV}$ in terms 
of effective constituent quark degrees of freedom. This approach has been
successful in describing various hadronic properties in terms 
of ``valence-quark degrees of freedom.'' The underlying idea is that 
at a low hadronic scale $\mu_0$, for example the properties of the nucleon 
can be modelled in terms of wave functions of valence $u$ and $d$ quarks, 
and similarly the properties of the $\pi^-$ in terms of the wave 
functions of valence $\bar u$ and $d$ quarks. It is an interesting task 
in itself to apply such a framework to the description of hadronic properties
like TMDs. This has been done in a variety of complementary approaches 
including chiral quark models~\cite{Weigel:1999pc} and generalizations~\cite{RuizArriola:2003bs}, spectator models (SPMs)~\cite{Jakob:1997wg,Gamberg:2007wm,Gamberg:2009uk,
Bacchetta:2008af,Lu:2004hu}, light-front constituent quark model (LFCQM)~\cite{Pasquini:2008ax,Pasquini:2010af,Lorce:2011dv,
Boffi:2009sh,Pasquini:2011tk,Pasquini:2014ppa,Lorce:2014hxa,Lorce:2016ugb} 
or bag models~\cite{Yuan:2003wk,Avakian:2008dz,Courtoy:2008vi,Courtoy:2008dn,
Avakian:2010br}. Phenomenological studies in the LFCQM showed that within a model accuracy of 20-30$\%$ a good description of SIDIS and unpolarized DY data can be
obtained \cite{Boffi:2009sh,Pasquini:2011tk,Pasquini:2014ppa}.

The goal of the present work is to study the spin and azimuthal asymmetries in the DY process with pions and polarized 
nucleons, and to present calculations for all twist-2 asymmetries.  
We use available phenomenological extractions of TMDs and 
calculations from two well-established 
constituent-quark-models (CQM), the LFCQM 
and the SPM. Other studies in models, perturbative QCD and lattice QCD 
of the pion-induced DY or relevant TMDs have been reported 
\cite{Noguera:2015iia,Engelhardt:2015xja,Lambertsen:2016wgj,Chang:2018pvk,
Bacchetta:2017vzh,Broniowski:2017gfp,Ceccopieri:2018nop}.

Several features distinguish our work from other studies. First, 
we use two CQM frameworks with diverse descriptions of the 
pion and nucleon structure. Second, we describe all leading-twist 
observables in pion-induced polarized DY entirely in the models.
Third, we supplement our studies with ``hybrid calculations'',  where we
use as much as possible information from phenomenological analyses, and 
only the 
Boer-Mulders function $h_{1,\pi}^{\perp a}$ is taken from models. 
Overall, we present up to four different calculations 
for each observable. This allows us to critically assess model dependence, 
and uncertainties in our approach. Where available the results are 
compared to the COMPASS DY data \cite{Aghasyan:2017jop}.

One key aspect in our study is the evolution of model results from 
the low hadronic scales to experimentally relevant scales. For that 
(i) knowledge of the low initial scale, and (ii) applicability of evolution 
equations at low scales are crucial. Both requirements are fulfilled in
the case of parton distribution functions which depend on one scale only, 
the renormalization scale~$\mu$.
First, the value of the initial quark model scale $\mu_0$ can be consistently
determined by evolving the fraction of nucleon momentum carried by valence 
quarks, $M_2^{\rm val}(\mu) = \sum_a \int dx\,x(f_1^q-f_1^{\bar q})(x,\mu)$,
known from parametrizations, using DGLAP evolution down to that scale
$\mu_0$ at which valence quarks carry the entire nucleon momentum, 
i.e.\ $M_2^{\rm val}(\mu_0) = 1$ \cite{Traini:1997jz}. 
Numerically it is $\mu_0\sim 0.5\,{\rm GeV}$. 
Second, works by the GRV and GRS groups on parametrizations of nucleon and 
pion unpolarized parton distribution functions show remarkable perturbative
stability between LO and NLO fits indicating applicability of DGLAP 
evolution down to initial scales as low as $\mu_0^2 = 0.26\,{\rm GeV}^2$ \cite{Gluck:1991ng,Gluck:1991ey,Gluck:1994uf,Gluck:1998xa,Gluck:1999xe}.

TMDs depend not only on the renormalization scale $\mu$ but also on 
the rapidity scale $\zeta$ \cite{Collins:2011zzd}. The theoretical and
phenomenological understanding of TMDs witnessed an incredible rate of 
developments in the recent years including NNLO and NNNLO calculations 
of the evolution kernel of unpolarized TMDs
\cite{Gehrmann:2014yya,Echevarria:2015byo,Echevarria:2015usa,Echevarria:2016scs,Li:2016ctv,Vladimirov:2016dll,Luo:2019hmp,Luo:2019szz,Ebert:2020yqt},
NLO calculations for the quark helicity distribution 
\cite{Gutierrez-Reyes:2017glx}, 
NLO \cite{Gutierrez-Reyes:2017glx} and NNLO \cite{Gutierrez-Reyes:2018iod}
calculations for transversity and pretzelosity, 
and NLO calculations for the Sivers function 
\cite{Ji:2006ub,Koike:2007dg,Sun:2013hua,Dai:2014ala,Scimemi:2019gge}.
Recently also the first non-trivial expression for the small-$b$
expansion of the pretzelosity distribution was derived \cite{Moos:2020wvd}.
However, in the context of quark model applications we face two challenges.
First, no rigorous (analog to the $\mu_0$-determination) criterion exists 
to fix the value of the initial rapidity scale $\zeta_0$ of quark models, 
though an educated guess may be $\zeta_0\sim\mu_0^2$.
Secondly, in the case of Collins-Soper-Sterman (CSS) or TMD evolution~\cite{Collins:1984kg,Collins:2011zzd},
no expertise is available analogous
to the GRV/GRS applications of DGLAP evolution starting from low hadronic scales.

In this situation in previous quark model studies, 
TMD evolution effects were often estimated approximately
\cite{Boffi:2009sh,Pasquini:2011tk,Pasquini:2014ppa} based on
an heuristic Gaussian Ansatz for transverse parton momenta with
energy dependent Gaussian widths. While providing a useful 
description of data on many processes including pion-induced Drell-Yan
\cite{Schweitzer:2010tt}, it is important to improve the simple Gaussian treatment in view of the recent progress in the TMD theory  
\cite{Gehrmann:2014yya,Echevarria:2015byo,Echevarria:2015usa,Echevarria:2016scs,Li:2016ctv,Vladimirov:2016dll,Luo:2019hmp,Luo:2019szz,Ebert:2020yqt,Gutierrez-Reyes:2017glx,Gutierrez-Reyes:2018iod,Ji:2006ub,Koike:2007dg,Sun:2013hua,Dai:2014ala,Scimemi:2019gge,Moos:2020wvd}.
We will therefore use TMD evolution~\cite{Collins:2011zzd} 
at  Next-to-Leading Logarithmic (NLL) precision 
to describe the transverse momentum dependence of the Drell-Yan process.
At present, application of TMD evolution at the low quark 
model scales below 1 GeV is not known. Therefore,  we shall proceed in two steps. We will evolve weighted
transverse moments of TMDs from the low initial scale $\mu_{0}^{2}$ to a scale of $Q_0^2=2.4\,{\rm GeV}^2$
where phenomenological information on transverse momentum dependence 
is available from TMD fits~\cite{Su:2014wpa, Kang:2014zza,Kang:2015msa, Bacchetta:2017gcc,Scimemi:2019cmh,Bacchetta:2019sam} of polarized and unpolarized SIDIS, DY and weak boson productions data. Then we use NLL TMD  evolution to evolve to the scales relevant in the 
COMPASS Drell-Yan measurements, i.e., $\la Q^2\ra=28\,{\rm GeV}^2$.
In this way we will be able to test the $x$-dependencies of the model TMDs while the $q_T$-dependencies of the DY observables are described on the basis of TMD fits. 

For completeness we remark that the importance of TMD evolution
for the description of pion-induced DY and the recent COMPASS
data was also studied in
Refs.~\cite{Wang:2017zym,Vladimirov:2019bfa,Li:2019uhj,Ceccopieri:2018nop,
Wang:2018naw,Wang:2018pmx}.

Our results serve several purposes. They help to interpret in their 
full complexity the first COMPASS data \cite{Aghasyan:2017jop} on the
pion-induced polarized DY process, and in this way deepen the understanding 
of the QCD description of deep-inelastic processes in terms of TMDs. 
They also provide quantitative tests of the application of  CQMs 
to the description of pion and nucleon structure.

\section{Drell-Yan process with pions and polarized protons}
 \label{sec-2}

In this section we briefly review the DY formalism, and 
provide the description of the DY structure functions in our approach.

\subsection{Structure functions}
\label{sec-2.1}

\begin{figure}[t!] 
\centering 
\includegraphics[height=6cm]{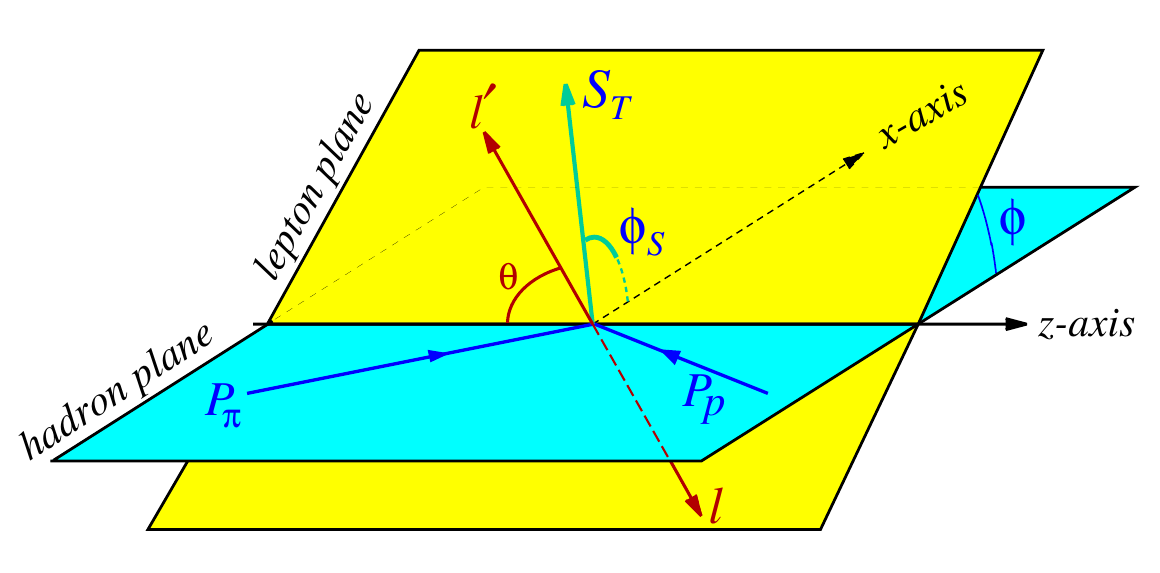}
\caption{\label{Fig-01:kinematics} 
    The DY process in the Collins-Soper frame where the pion
    and the proton come in with different momenta $P_\pi$, $P_p$, but each
    carries the same transverse momentum $\frac12\,\qT$, and 
    the produced lepton pair is at rest. The angle $\phi$ describes the 
    inclination of the leptonic frame with respect to the hadronic plane, 
    and $\phi_S$ is the azimuthal angle of the transverse-spin vector 
    of the proton.
    }
\end{figure}

In the tree-level description a dilepton $l,\,l^\prime$ is produced from
the annihilation of a quark and antiquark carrying the fractions $x_\pi$, 
$x_p$ of the longitudinal momenta of respectively the pion and the proton. 
The process is shown in the Collins-Soper frame in Fig.~\ref{Fig-01:kinematics}.
In the case of pions colliding with polarized protons the DY cross 
section is described in terms of six structure functions \cite{Arnold:2008kf},  
\ba
   F_{UU}^1 &=& \phantom{-}
   {\cal C}  \bigg[f_{1,\pi}^{\bar a} \; f_{1,p}^{a}\;\bigg], \nonumber \\
   F_{UU}^{\cos 2\phi} &=& \phantom{-}
   {\cal C} \bigg[ \frac{2(\hhat \cdot \vkTpi)(\hhat \cdot \vkTN) 
       - \vkTpi \cdot \vkTN}{M_\pi \; M_p} \; 
     h_{1,\pi}^{\perp \bar a} \; h_{1,p}^{\perp a} \; \bigg],  \nonumber \\
   F_{UL}^{\sin 2\phi} &=& 
   -  {\cal C} \bigg[ \frac{2(\hhat \cdot \vkTpi)(\hhat \cdot \vkTN) 
       - \vkTpi \cdot \vkTN}{M_\pi \; M_p} \; 
     h_{1,\pi}^{\perp\bar a}\; h_{1L, p}^{\perp a} \; \bigg],  \nonumber \\
   F_{UT}^{\sin \phi_S} &=& \phantom{-}
   {\cal C} \bigg[ \frac{\hhat \cdot \vkTN}{M_p} \; 
     f_{1,\pi}^{\bar a} \; f_{1T, p}^{\perp a} \; \bigg],  \nonumber \\
   F_{UT}^{\sin (2\phi - \phi_S)} &=& 
   -  {\cal C}  \bigg[ \frac{\hhat \cdot \vkTpi}{M_\pi} \; 
     h_{1,\pi}^{\perp\bar a} \;h_{1, p}^{a} \; \bigg],  \nonumber\\
   F_{UT}^{\sin (2\phi + \phi_S)} &=& 
   -  {\cal C}  \bigg[ \frac{2(\hhat \cdot \vkTN) 
       \big[2(\hhat \cdot \vkTpi)(\hhat \cdot \vkTN) 
       - \vkTpi \cdot \vkTN \big] 
       - \vkTN^2(\hhat \cdot \vkTpi)}{2 \; M_\pi \; M_p^2} \,
     h_{1, \pi}^{\perp\bar a}\, h_{1T, p}^{\perp a} \; \bigg]. \quad \quad \label{Eq:SFs} 
\ea
The subscripts indicate the hadron polarization which can be unpolarized $U$ 
(pions, protons), longitudinally $L$, or transversely $T$ polarized (protons). 
The azimuthal angles $\phi$, $\phi_S$ are defined in
Fig.~\ref{Fig-01:kinematics}, where the unit vector 
$\hhat=\qT/q_T$ points along the $x$-axis. Notice that in 
the Collins-Soper frame the dilepton is at rest, and 
each incoming hadron carries the transverse momentum 
$\qT/2$, see Fig.~\ref{Fig-01:kinematics}.
The convolution integrals in Eq.~\eqref{Eq:SFs}
are defined  as~\cite{Arnold:2008kf} 
\be
 \label{Eq:convolution-integral} 
    {\cal C}\bigl[\omega \, f_{\pi}^{\bar a} \, f_p^{a}\bigr] = \frac{1}{N_c} \; 
    \sum_a e_a^2\int d^2\kTpi \, d^2\kTN \, \delta^{(2)}(\qT- \kTpi-\kTN) \,
    \omega \, f_{\pi}^{\bar a}(x_\pi,\kTpi^2)f_p^{a}(x_p,\kTN^2)\,,
\ee  
where $\omega$, which is a function of the transverse momenta $\kTpi$, $\kTN$
and $\qT$, projects out the corresponding azimuthal angular dependence.
The sum over $a=u,\,\bar{u},\,d,\,\bar{d},\,\dots$ includes the active flavors. 

This partonic interpretation of DY is based on a TMD factorization~\cite{Collins:1984kg,Collins:2011zzd} and applies to the region $q_T\ll Q$. 
The TMDs depend on renormalization and rapidity scales which are not
indicated for brevity in (\ref{Eq:SFs}) and (\ref{Eq:convolution-integral}), and will be discussed in section~\ref{sec-2.2}.
The focus of our work is on asymmetries of the kind
\be
 \label{Eq:Asym.def} 
    A_{XY}^{\rm weight}(x_\pi, x_p , q_T, Q^2)= 
    \frac{F_{XY}^{\rm weight}(x_\pi, x_p , q_T, Q^2)}{F_{UU}^1(x_\pi, x_p , q_T, Q^2)},
\ee 
where various types of higher order corrections tend to largely 
cancel out \cite{Ratcliffe:1982yj,Weber:1991wd,Vogelsang:1992jn,
Contogouris:1994ws,Gehrmann:1997pi,Bunce:2000uv,Shimizu:2005fp}.

The $Q^2$ dependence of the structure functions and asymmetries will often
not be explicitly indicated for brevity. 
In the following we will display results for the asymmetries as functions 
of one of the variables $x_\pi$, $x_p$, $q_T$. It is then understood that 
the structure functions are integrated over the other variables 
within the acceptance of the experiment, keeping in mind that $x_\pi, \, x_p$
are connected to each other by $x_\pi \, x_p = Q^2/s$, where  $s$ is the center of mass energy squared.

\subsection{QCD evolution of Drell-Yan structure functions}
\label{sec-2.2} 

The basis for the evolution  are TMD factorization theorems~\cite{Collins:1981uk,Collins:1984kg,Qiu:1991pp,Ji:2004wu,Ji:2006br,Ji:2006vf,Collins:2011zzd,Aybat:2011zv,Ma:2013aca,Collins:2014jpa,Collins:2016hqq,Scimemi:2018xaf}
which constrain the operator definition 
and define the QCD evolution of TMDs. 
Here we will adopt the CSS framework 
and use the TMD evolution formalism starting from a fixed scale $Q_0$
\cite{Collins:2014jpa} in the structure functions from  Eqs.~\eqref{Eq:SFs}. 
 
The evolution of TMDs is a double-scale problem, and 
can be implemented in momentum space or impact-parameter space with examples for both approaches in the literature 
\cite{Collins:1981va,Aybat:2011zv,Angeles-Martinez:2015sea,Scimemi:2019cmh,Bacchetta:2019sam,Ebert:2020dfc}.
In our work we choose to implement the TMD evolution
in the impact-parameter space with ${\bm b_{T}}$ the 
Fourier-conjugate variable to $\kTh$
where index $h = \pi$ or $p$ refers to pion or nucleon. 
The TMDs in the impact-parameter
space are generically given by $\tilde f(x_h,{ b_{T}},\mu,\zeta)$ where
$\mu\sim Q$ is the ``standard'' renormalization scale for ultraviolet logarithms, 
and $\zeta\sim Q^2$ is the rapidity renormalization scale.
In principle one can solve TMD evolution equations starting from some 
initial scale $Q_0$ without employing operator product expansion at low 
${\bm b_{T}}$,   Ref.~\cite{Collins:2014jpa}. 
The TMD at this initial scale is then $f(x_h,{b_{T}},Q_0,Q_0^2)$. 
In this formulation the unpolarized structure function is similar to 
parton model result and is 
expressed as~\cite{Collins:2014jpa}
\begin{align}
\label{FUUbspace} F_{UU}^1(x_\pi,x_p,q_T,Q^2) &= \frac{1}{N_c} \sum_a e_a^2 {\cal H}^{(DY)}(Q,\mu_Q)\int \frac{b_{T} d b_{T}}{2\pi} J_0 (q_T b_{T}) \nonumber \\ 
   & \times  f_{1,\pi}^{\bar a}(x_\pi,{b_{T}},Q_0,Q_0^2) \tilde f_{1,p}^a(x_p,{b_{T}},Q_0,Q_0^2)\, e^{-S(b_{T},Q_0,Q,\mu_Q)}\; ,
\end{align}
where the factor $S(b_T,Q_0,\mu_Q)$ contains important effects of gluon
radiation with $S(b_T,Q_0,Q_0) = 0$ by construction~\cite{Collins:2014jpa}. The hard factor ${\cal H}(Q,\mu_Q)$ is~\cite{Collins:2017oxh}
\begin{align}
    {\cal H}^{(DY)}(Q,\mu_Q) = 1 + \frac{\alpha_s(\mu_Q)}{2 \pi} C_F \left( 3 \ln \left( \frac{Q^2}{\mu_Q^2}\right) - \ln^2 \left( \frac{Q^2}{\mu_Q^2}\right) +\frac{7\pi^2}{6} - 8\right)\; + {\cal O} (\alpha_s^2) ,
    \label{eq:hard}
\end{align}
where $C_F=4/3$ and $\alpha_s$ is the strong coupling constant.

One can parametrize TMDs at initial scale $Q_0$ as 
\be
 \label{Ansatzbspace} 
     \tilde f_{1,p}^{a}(
     x_p,{ b_{T}},Q_0,Q_0^2) = f_{1,p}^{a} (
     x_p,Q_0) \;
     e^{-\frac14 b_{T}^2
     \avkTN_{f_{1,p}}} \,,
\ee
\be
 \label{Ansatzbspace1} 
     \tilde f_{1,\pi}^{a}(x_\pi,{ b_{T}},Q_0,Q_0^2) = f_{1,\pi}^{a} (x_\pi,Q_0) \; e^{-\frac14 b_{T}^2\avkTpi_{f_{1,\pi}}} \,,
\ee
where $x$-dependent functions correspond to collinear distributions and the exponential factors are ``primordial shapes'' of TMDs at the initial scale. This particular dependence is often used in phenomenology \cite{DAlesio:2004eso,Schweitzer:2010tt},
corresponds to the Gaussian Ansatz and is supported in models
\cite{Avakian:2010br,Efremov:2010mt,Pasquini:2011tk,Pasquini:2014ppa,Schweitzer:2012hh}. The average widths of TMDs may be flavor-
and $x$-dependent and will be taken from
phenomenological parametrizations at $Q_0^2$. 

  Based on the $b_T$ space formalism given in Ref.~\cite{Boer:2011xd} we  write down the rest of the twist-2 structure functions. We use the convenient notation from Ref.~\cite{Bacchetta:2019qkv}, 
\begin{align}
     {\cal B}_n[\tilde f_{\pi}\; \tilde f_{p}] & \equiv  \frac{1}{N_c} 
     \sum_a e_a^2 {\cal H}^{(DY)}(Q,\mu_Q) \int_0^\infty \frac{d b_T\, b_T}{2\pi}\; b_T^n \, J_n(q_T b_T)
     \nn &\hskip -1cm \times  \tilde f_{\pi}^{\bar a}(x_\pi,{b_{T}},Q_0,Q_0^2) 
     \; \tilde f_{p}^{ a}(x_p,{b_{T}},Q_0,Q_0^2) \; e^{-S(b_{T},Q_0,Q,\mu_Q)}\, ,
    \label{Eq:convBess}
\end{align}
 which leads to the  following expressions for  the   twist-2 structure functions,
\begin{align}
{F}_{UU}^1(x_\pi,x_p,q_T,Q^2) &=
   \phantom{-}   {\cal B}_0[\tilde f_{1, \pi}\; \tilde f_{1,p}] \; ,
    \label{e:upb} \\
{F}_{UU}^{\cos 2\phi}(x_\pi,x_p,q_T,Q^2) &= 
  \phantom{-}   M_\pi M_p \; {\cal B}_2[\tilde h_{1,\pi}^{\perp(1)}\; \tilde h_{1,p}^{\perp(1)}]\; ,
\label{e:bmb}   \\
{F}_{UL}^{\sin 2\phi}(x_\pi,x_p,q_T,Q^2)&=
   - M_\pi  M_p \; {\cal B}_2[\tilde h_{1,\pi}^{\perp(1)}\; \tilde  h_{1L,p}^{\perp(1)}]\; ,
   \label{e:kotzb} \\
{F}_{UT}^{\sin \phi_S}(x_\pi,x_p,q_T,Q^2) &= 
  \phantom{-} M_p \;{\cal B}_1[\tilde f_{1,\pi}\; \tilde  f_{1T,p}^{\perp(1)}]\; ,
  \label{e:Sivb}\\
{F}_{UT}^{\sin (2\phi -\phi_S)}(x_\pi,x_p,q_T,Q^2) &=
    -M_\pi \;{\cal B}_1[\tilde h_{1,\pi}^{\perp(1)}\; \tilde  h_{1,p}]
    \label{e:bmtb}\; ,\\
{F}_{UT}^{\sin (2\phi + \phi_S)}(x_\pi,x_p,q_T,Q^2) &=
    -\frac{M_\pi M_p^2}{4} \; {\cal B}_3[\tilde h_{1,\pi}^{\perp(1)}\; \tilde  h_{1T,p}^{\perp(2)}]
\; ,\label{Eq:sfsb}
\end{align}
where the $b_T$ space TMD moments~\cite{Boer:2011xd} are
\begin{align}
    \tilde f^{(n)}(x_h,b_T,Q,Q^2) &=(-1)^n  n! \left(\frac{2}{M_h^2}\frac{\partial}{\partial b_T^2}\right)^n 
    \tilde f(x_h,b_T,Q,Q^2) \; .
\end{align}
 These moments have the important feature,
 
\begin{align}
    \lim_{b_T \to 0} \tilde f^{(n)}(x_h,b_T,Q,Q^2) = f^{(n)}(x_h,Q)\; ,
\end{align}
where $f^{(n)}$ are   conventional transverse moments of TMDs~\cite{Mulders:1995dh}  defined as
\be
       f^{(n)}(x_h,Q)=\int\di^2\kTh\,\left(\frac{\kTh^2}{2M_h^2}\right)^n\,f(x_h,\kTh^2,Q,Q^2)\, ,
       \label{e:TMDmoment}
\ee
and $h=\pi, p$ corresponds to pion and proton TMDs, respectively.
The evolution factor $S(b_T, Q_0, \mu_Q)$   in Eq.~\eqref{Eq:convBess},  
which results from  
solving the CSS evolution equation and the renormalization group
equations for the rapidity dependence of the TMDs
and for the soft factor~\cite{Aybat:2011zv,Collins:2011zzd}, is given by

\begin{eqnarray}
S(b_T, Q_0, Q, \mu_Q) =  -\tilde K(b_T,Q_0) \ln\frac{Q^2}{Q_0^2}  +
 \int_{Q_0}^{\mu_Q}\frac{d\bar\mu}{\bar\mu}\left[\gamma_K(\alpha_s(\bar\mu))\ln\frac{Q^2}{\bar\mu^2}-2 \gamma_i(\alpha_s(\bar\mu);1)\right]\, ,
 \label{e:FF_ansatz}
\end{eqnarray}

\noindent
where $\tilde K$ is the Collins-Soper evolution kernel, and the anomalous dimensions are $\gamma_i(\alpha_s(\bar\mu);1)$ and $\gamma_K(\alpha_s(\bar\mu))$~\cite{Collins:2014jpa}.

Since the integral in Eq.~\eqref{Eq:convBess} extends over all $b_T$,
one cannot avoid using $\tilde K$ in the 
CSS evolution factor~\eqref{e:FF_ansatz}  in the non-perturbative large $b_T$ region.
In order to combine the perturbative and non-perturbative regions,
we use the $\bstarsc$ prescription~\cite{Collins:1984kg}, namely,  
 \begin{align}
\label{eq:bstar}
  \bstarsc = \frac{b_T }{\sqrt{ 1 + b_T^2/\bmax^2}},
\end{align}
which introduces a smooth upper cutoff $b_{\rm max}$  in the transverse distance.

 Then, the perturbative part of 
$\tilde{K}$ is defined by replacing $b_T$ by $\bstarsc$ and the non-perturbative part is defined by the difference  
$\tilde{K}(\bstarsc,\mu)-\tilde{K}(b_T,\mu)=g_K(b_T;\bmax)~$\cite{Collins:2014jpa}.
Furthermore  to combine the perturbative and non-perturbative regions
using the fixed scale evolution, it is optimal  to use the renormalization group running scheme for  $\tilde K$ in Eq.~\eqref{e:FF_ansatz}, evolved from the fixed scale $Q_0$, i.e.

\bea
\tilde K(b_T,Q_0)=\tilde K(\bstarsc,\mubstar) -\int_{\mubstar}^{Q_0}\frac{d\bar\mu}{\bar\mu}\gamma_K(\alpha_s(\bar\mu)) -g_K(b_T;b_{\rm max}) \, ,
\eea
where $\mubstar$ is now chosen to become a hard scale,  
\begin{eqnarray}
\mubstar \equiv \frac{C_1}{b_*}\, .
\end{eqnarray}
Now  Eq.~\eqref{e:FF_ansatz} reads~\cite{Collins:2014jpa}
\begin{align}
S(b_T,\bstarsc, Q_0, Q, \mu_Q) &=  \left( g_K(b_T;b_{\rm max})- \tilde K(\bstarsc;\mubstar)+
  \int_{\mubstar}^{Q_0}\frac{d\bar\mu}{\bar\mu}\gamma_K(\alpha_s(\bar\mu)) \right)\ln\frac{Q^2}{Q_0^2}  
 \nonumber \\
 + & \int_{Q_0}^{\mu_Q}\frac{d\bar\mu}{\bar\mu}\left[\gamma_K(\alpha_s(\bar\mu))\ln\frac{\mu_Q^2}{\bar\mu^2}-2 \gamma_i(\alpha_s(\bar\mu);1)\right]\, .
 \label{e:final}
\end{align}
The anomalous  dimensions  can be expanded as
perturbative series,  
$\gamma_i =\sum_{n=1}^\infty \gamma_i^{(n)} \left(\alpha_s/\pi\right)^n$, and $\gamma_K =\sum_{n=1}^\infty \gamma_K^{(n)} \left(\alpha_s/\pi\right)^n$.  In our calculations we employ  them to NLL accuracy:
$\gamma_K^{(1)}$, $\gamma_K^{(2)}$ and $\gamma_i^{(1)}$.
They are spin-independent~\cite{Collins:1984kg,Qiu:2000ga,Landry:2002ix,Moch:2005id,Kang:2011mr,Aybat:2011zv,Echevarria:2012pw,Grozin:2014hna,Collins:2014jpa},  and given by
\bea
\gamma_K^{(1)} =2 C_F,
\quad
\gamma_K^{(2)} ={C_F}\left[C_A\left(\frac{67}{18}-\frac{\pi^2}{6}\right)-\frac{10}{9}T_F \, n_f\right],
\quad
\gamma_i^{(1)} = \frac{3}{2} C_F
\, ,
\eea
where $C_F=4/3$, $C_A=3$, $T_F=1/2$ and $n_f$ is the number of active flavors.
The NLL two-loop contribution for $\tilde K$~\cite{Collins:2017oxh,Echevarria:2020hpy}, valid  at small values of $b_T$, is
\begin{align}
 \tilde K (b_*;\mubstar) &= -2 C_F \frac{\alpha_s(\mubstar)}{\pi} \ln \left(\frac{b_* \mu_{b_*}}{2 e^{-\gamma_E}}\right) 
 +
 \frac{C_F}{2}\left(\frac{\alpha_s(\mubstar)}{\pi}\right)^2\Bigg[ \left( \frac{2}{3} n_f - \frac{11}{3}C_A\right) \ln^2 \left(\frac{b_* \mu_{b_*}}{2 e^{-\gamma_E}}\right)
 \nonumber \\
 &+
 \left( -\frac{67}{9} C_A + \frac{\pi^2}{3}C_A + \frac{10}{9} n_f\right) \ln \left(\frac{b_* \mu_{b_*}}{2 e^{-\gamma_E}}\right) +
   \left(\frac{7}{2}\zeta_3-\frac{101}{27}\right) C_A + \frac{28}{27} T_F n_f 
 \Bigg] \, ,
 \label{e:CS_kernel0}   
\end{align}
so that for $C_1 =2 e^{-\gamma_E}$, one finds
\begin{eqnarray}
\tilde K (b_*;\mubstar) &=& \frac{C_F}{2} \left(\frac{\alpha_s(\mubstar)}{\pi}\right)^2  \left[\left(\frac{7}{2}\zeta_3-\frac{101}{27}\right) C_A + \frac{28}{27} T_F n_f 
 \right] \, .
 \label{e:CS_kernel}
\end{eqnarray}
 We will numerically calculate  the integral in Eq.~\eqref{e:FF_ansatz} using the two-loop result for the strong coupling constant, tuned to the world average~\cite{Bethke:2012jm} $\alpha_s(M_Z)= 0.118$ as in the CTEQ analysis~\cite{Hou:2019efy}.

Furthermore, we adopt the functional form of $g_K(b_T;b_{\rm max})$ given by Collins and Rogers~\cite{Collins:2014jpa},
\bea
g_K(b_T,\bmax)&=&g_0(\bmax)\left(1-\exp\left[-\frac{C_F\alpha_s(\mubstar)b_T^2}{\pi g_0(\bmax)\bmax^2}\right]\right),
\eea
 which interpolates smoothly between the small and large-$b_T$ regions, where
at small $b_T$ it approximates a power series in $b_T^2$, while at large $b_T$ the resulting value of $\tilde K$ goes to a constant~\cite{Collins:2014jpa}. We choose  $g_0(\bmax)=0.84$ and $b_{max} = 1\,{\rm GeV}^{-1}$ to match the non-perturbative behavior of $g_K$ used in Refs.~\cite{Kang:2014zza,Kang:2015msa} to describe the  polarized SIDIS data and in Ref.~\cite{Su:2014wpa} to describe unpolarized SIDIS, DY and weak boson production data. 
The scale $\mu_Q$ is usually chosen such that $\mu_Q= C_2 Q $. In the following we will use $C_2 = 1$ and $C_1 = 2 e^{-\gamma_E}$. These choices allow one to optimize the accuracy of the perturbative expansion calculations in Eqs.~(\ref{eq:hard}) and (\ref{e:CS_kernel0})~\cite{Collins:2017oxh}.  
\subsection{
Input for TMDs and choice of the initial scale \boldmath{$Q_0$}}
\label{sec-2.3}
We will utilize the following parametrizations~\cite{Bastami:2018xqd}
for TMDs at the initial scale $Q_0$:
\ba
f_{h}^a(x_h,\kTh,Q_0,Q_0^2) &=& 
f_{h}^a(x_h,Q_0) \; \frac{e^{- \kTh^2 / \avkTh_{f_{h}}  }}{\pi \; \avkTh_{f_{h}} }\; , 
\quad \hspace{16mm}
f_{h}^a = f_{1, p}^a, \; f_{1, \pi}^a,\; h_{1, p}^a, \nonumber\\
f_{h}^{a}(x_h,\kTh,Q_0,Q_0^2) &=&  
f_{h}^{(1) a}(x_h,Q_0) \, \frac{2 M_h^2}{\pi \avkTh_{f_{h}}^2} e^{-\kTh^2/{\avkTh_{f_{h}}}}, 
\quad
f_h^a = f_{1T, p}^{\perp a},\;h_{1, p}^{\perp a},\;h_{1, \pi}^{\perp a},\;h_{1L, p}^{\perp a},\nonumber\\
f_{h}^{a}(x_h,\kTh,Q_0,Q_0^2) &=&  
f_{h}^{(2) a}(x_h,Q_0) \, \frac{2 M_h^4}{\pi \avkTh_{f_{h}}^3} e^{-\kTh^2/{\avkTh_{f_{h}}}}, 
\quad
f_h^a = h_{1T, p}^{\perp a}, \label{Eq:Gauss}
\ea
where transverse moments of TMDs are defined in Eq.~\eqref{e:TMDmoment}. Parametrizations from Eqs.~\eqref{Eq:Gauss} are often used in phenomenology to describe polarized SIDIS and DY data. 
These parametrizations correspond to the following $b_T$-space expressions
\ba
\tilde f_{h}^a(x_h,b_T,Q_0,Q_0^2) &=& 
f_{h}^a(x_h,Q_0) \;  e^{- \frac{1}{4} \, b_T^2 \avkTh_{f_{h}}} \; , 
\quad \hspace{2mm}
\tilde f_{h}^{a} = \tilde f_{1, p}^a, \; \tilde f_{1, \pi}^a,\; \tilde h_{1, p}^a, \nonumber\\
\tilde f_{h}^{(1) a}(x_h,b_T,Q_0,Q_0^2) &=&  
f_{h}^{(1) a}(x_h,Q_0) \,  e^{-\frac{1}{4} \, b_T^2 {\avkTh_{f_{h}}}}, 
\quad
\tilde f_h^{(1)a} = \tilde f_{1T, p}^{\perp (1) a},\;\tilde h_{1, p}^{\perp (1) a},\;\tilde h_{1, \pi}^{\perp (1) a},\;\tilde h_{1L, p}^{\perp (1) a},\nonumber\\
\tilde f_{h}^{(2) a}(x_h,b_T,Q_0,Q_0^2) &=&  
f_{h}^{(2) a}(x_h,Q_0) \,  e^{-\frac{1}{4} \, b_T^2{\avkTh_{f_{h}}}}, 
\quad
\tilde f_h^{(2)a} = \tilde h_{1T, p}^{\perp (2) a}, \label{Eq:Gaussb}
\ea
where the collinear functions are the same as in Eqs.~\eqref{Eq:Gauss}.

Using Eqs.~\eqref{Eq:Gauss}  or Eqs.~\eqref{Eq:Gaussb} one obtains for the convolution integrals 
in Eqs.~(\ref{Eq:SFs}) or Eqs.~\eqref{e:upb}--\eqref{Eq:sfsb} the following results at the initial scale $Q_0$, 
\ba 
     F_{UU}^1(x_{\pi},x_p,q_T,Q_0^2)
     &=& \phantom{-} \frac{1}{N_c} \sum_a e_a^2 \; 
     \;f_{1,\pi}^{\bar{a}} (x_{\pi},Q_0)\;f_{1,p}^a(x_p,Q_0)
     \;\frac{e^{-q_T^2/\avqT}}{\pi \; \avqT}, \nonumber\\
     F_{UT}^{\sin\phi_S}(x_{\pi},x_p,q_T,Q_0^2)
     &=& \phantom{-} \frac{1}{N_c} \sum_a e_a^2 \; 
     f_{1,\pi}^{\bar{a}} (x_{\pi},Q_0) \; f_{1T,p}^{\perp (1)a}(x_p,Q_0)
     \; 2 M_p \; \frac{q_T }{\avqT} \; 
     \frac{e^{-q_T^2/\avqT}}{\pi \; \avqT}, \nonumber\\
     F_{UT}^{\sin(2\phi-\phi_S)}(x_{\pi},x_p,q_T,Q_0^2)
     &=& -  \frac{1}{N_c} \sum_a e_a^2 \; 
     h_{1,\pi}^{\perp (1) {\bar{a}}} (x_{\pi},Q_0) \; h_{1,p}^a(x_p,Q_0)
     \; 2 M_{\pi} \; \frac{q_T }{\avqT} \;
     \frac{e^{-q_T^2/\avqT}}{\pi \; \avqT}, \nonumber\\
     F_{UU}^{\cos 2\phi}(x_{\pi},x_p,q_T,Q_0^2)
     &=& \phantom{-} \frac{1}{N_c} \sum_a e_a^2 \; 
     h_{1,\pi}^{\perp (1) {\bar{a}}} (x_{\pi},Q_0) \; h_{1,p}^{\perp (1) a}(x_p,Q_0)\;
     4 M_{\pi}M_p \;  \frac{q_T^2 }{\avqT^2} \; 
     \frac{e^{-q_T^2/\avqT}}{\pi \; \avqT},\nonumber\\
     F_{UL}^{\sin 2\phi}(x_{\pi},x_p,q_T,Q_0^2)
     &=& - \frac{1}{N_c}  \sum_a e_a^2 \; 
     h_{1,\pi}^{\perp (1){\bar{a}}} (x_{\pi},Q_0) \; h_{1L,p}^{\perp (1)a}(x_p,Q_0)
     \; 4 M_{\pi}M_p\; \frac{q_T^2 }{\avqT^2} \; 
     \frac{e^{-q_T^2/\avqT}}{\pi \; \avqT}, \nonumber\\
     F_{UT}^{\sin(2\phi+\phi_S)}(x_{\pi},x_p,q_T,Q_0^2)
     &=& -  \frac{1}{N_c} \sum_a e_a^2 \; 
     h_{1,\pi}^{\perp (1){\bar{a}}} (x_{\pi},Q_0) \; h_{1T,p}^{\perp (2) a}(x_p,Q_0)
     \; 2 M_{\pi}M_p^2\;  
     \frac{q_T^3 }{\avqT^3} \; 
     \frac{e^{-q_T^2/\avqT}}{\pi \; \avqT}, \nonumber\\   \label{Eq:all-Gauss} 
\ea
where the index $a=u,\,\bar{u},\,d,\,\bar{d},\,\dots$ 
and the mean square transverse momenta $\avqT$ are defined in each case as 
the sum of the mean square transverse momenta of the corresponding TMDs; that is in  (\ref{Eq:all-Gauss})
in the first equation $\avqT = \avkTpi_{f_{1,\pi}}+\avkTN_{f_{1,p}}$, 
in the second equation $\avqT = \avkTpi_{f_{1,\pi}}+\avkTN_{f^\perp_{1T,p}}$, etc.

In our study we will use the corresponding 
transverse moments and Gaussian widths from TMD extractions
that we will take at the initial scale
which corresponds to the $\la Q^2\ra$ in the HERMES experiment. We will therefore use $Q_0^2 = 2.4$ GeV$^2$ as our initial scale of TMD evolution in Eqs.~\eqref{Eq:Gaussb} for parametrization of TMDs 
and in Eqs.~\eqref{e:upb}-~\eqref{Eq:sfsb} for structure functions 
that we will evolve to the scale of the COMPASS Drell-Yan measurement.

The scale $Q_0^2=2.4\,{\rm GeV}^2$ is convenient because parametrizations of
many TMDs are available at this scale, and there is a great deal of
expertise how to implement CSS evolution starting from $Q_0$.
However, results from quark models refer to a lower hadronic 
scale $\mu_0 \sim 0.5\,{\rm GeV} < Q_0$. Presently it is not 
known how to implement CSS evolution at such low scales (see section~\ref{sec:intro}). Thus, the evolution of model results from $\mu_0$ 
to the initial CSS scale $Q_0$ chosen in this work, is 
regarded as a part of the modelling. It will be described 
in detail in section~\ref{sec-2.5}.

\subsection{TMDs extracted from experimental data}
\label{sec-2.4}

\begin{figure}[b!] 
\centering 
\includegraphics[height=5.3cm]{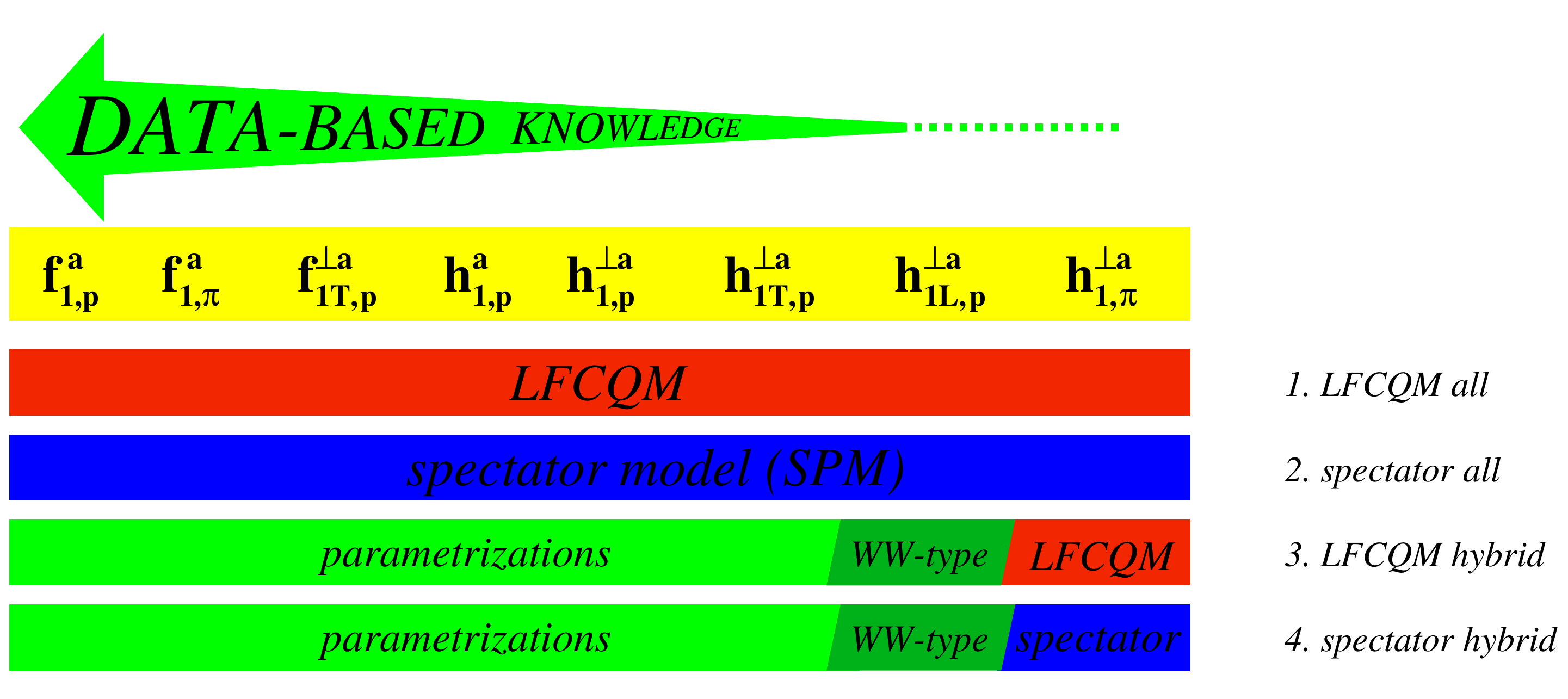}
\caption{\label{Fig-02:knowledge-TMDs} 
    TMDs entering the pion-induced polarized DY 
    process at leading twist in the order from 
    the phenomenologically best to least known, 
    and the approaches used in this work, see text.}
\end{figure}

In order to compute leading-twist structure functions in pion-induced
DY the knowledge of the proton and pion TMDs $f_{1,p}^a$,
$f_{1,\pi}^a$, $f_{1T,p}^{\perp a}$, $h_{1,p}^a$, $h_{1,p}^{\perp a}$,
$h_{1T,p}^{\perp a}$, $h_{1L,p}^{\perp a}$, $h_{1,\pi}^{\perp a}$
is required, which we list here  in the order from the best to the
least known TMD, see Fig.~\ref{Fig-02:knowledge-TMDs} for an overview. 
While such a classification is to some extent subjective, it is evident that the collinear proton distributions $f_{1,p}^a(x_p)$ 
are the best known
\cite{Gluck:1998xa,Martin:2009iq,Harland-Lang:2014zoa,Dulat:2015mca}
thanks to DIS, DY and other data. We will utilize the MSTW extraction 
of $f_{1,p}^a(x_p)$ \cite{Martin:2009iq} for comparison with models 
and our calculations. 
The unpolarized TMDs $f_{1,p}^a(x_p,\kTN)$ have been studied 
and much progress was achieved in incorporating effects of QCD evolution
\cite{Landry:2002ix,Anselmino:2013lza,Signori:2013mda,Bacchetta:2019sam,Scimemi:2019cmh}
which are taken into consideration approximately in our approach
as described in section~\ref{sec-2.2}. For the collinear pion 
distribution $f_{1,\pi}^a$, listed next in Fig.~\ref{Fig-02:knowledge-TMDs}, 
many extractions are available
\cite{Gluck:1991ey,Sutton:1991ay,Gluck:1999xe,Aicher:2010cb,Barry:2018ort,Novikov:2020snp}. We will use the MRSS fits \cite{Sutton:1991ay}.

One of the most prominent TMDs, the Sivers distribution $f_{1T,p}^{\perp a}$
was extracted from HERMES, COMPASS, and JLab SIDIS data by several groups 
with consistent results
\cite{Anselmino:2010bs,Anselmino:2005ea,Anselmino:2005an,Collins:2005ie,Vogelsang:2005cs,Anselmino:2008sga,Bacchetta:2011gx,Anselmino:2011gs,Echevarria:2014xaa,Cammarota:2020qcw,Bacchetta:2020gko}. We will use the extractions of 
Ref.~\cite{Anselmino:2011gs} labelled as ``Torino'' and
Ref.~\cite{Cammarota:2020qcw} labelled as ``JAM20''. 

The transversity distribution, $h_{1,p}^a$, plays a crucial role in
understanding the nucleon spin structure. It is predicted to generate a transverse single spin
asymmetry in SIDIS coupling to the Collins fragmentation function
\cite{Collins:1992kk}, which is also responsible for an azimuthal asymmetry  in $e^+e^-$ annihilation into hadron pairs. 
We will use the ``Torino'' parametrizations of $h_{1,p}^a$
from a  global QCD analysis of SIDIS and $e^+e^-$ data
\cite{Anselmino:2013vqa} to be compared with model predictions, 
and the ``JAM20'' fit from a global QCD analysis of SIDIS, DY,
$e^+e^-$, and proton-proton data \cite{Cammarota:2020qcw} for 
comparisons and calculations. 

The proton Boer-Mulders function $h_{1,p}^{\perp a}$
extracted from HERMES, COMPASS and DY data
in Ref.~\cite{Barone:2009hw} will be used with the label
``BMP10.'' The extraction of $h_{1,p}^{\perp a}$
\cite{Barone:2009hw} is less certain, because 
in SIDIS it requires model-dependent corrections 
for sizable twist-4 contamination (Cahn effect).

The so-called pretzelosity function $h_{1T,p}^{\perp a}$  was extracted 
in Ref.~\cite{Lefky:2014eia}. We will label $h_{1T,p}^{\perp a}$  from Ref.~\cite{Lefky:2014eia} as ``LP15''. 
Notice that large errors on extracted $h_{1T,p}^{\perp a}$ were 
reported in Ref.~\cite{Lefky:2014eia}. 
This is the least known proton TMD for which an extraction has been attempted.

Only the Kotzinian-Mulders distribution $h_{1L,p}^{\perp a}$ has not 
yet been extracted. It was found that the data related to this TMD \cite{Airapetian:1999tv,Jawalkar:2017ube,Parsamyan:2018evv} 
are compatible with the WW-type approximation
\cite{Bastami:2018xqd} which we will use to approximate
$h_{1L,p}^{\perp a}$  based on $h_{1,p}^a$ 
from ``Torino'' \cite{Anselmino:2013vqa} and ``JAM20'' \cite{Cammarota:2020qcw} fits.

Finally, the pion Boer-Mulders function  $h_{1,\pi}^{\perp a}$ 
is the least known of the TMDs needed to describe the pion-proton
DY process at leading twist. No extractions  
are currently available for this TMD.

\subsection{TMDs from models}
\label{sec-2.5} 

\begin{figure}[b]
\centering
	\begin{tabular}{cccc}

		\rotatebox[origin=c]{90}{$x \, f_{1,\, \pi^-}^{\bar u}$} &\vspace{-.5mm} \hspace{-3mm}\raisebox{-.5\height}{\includegraphics[width=0.33\textwidth]{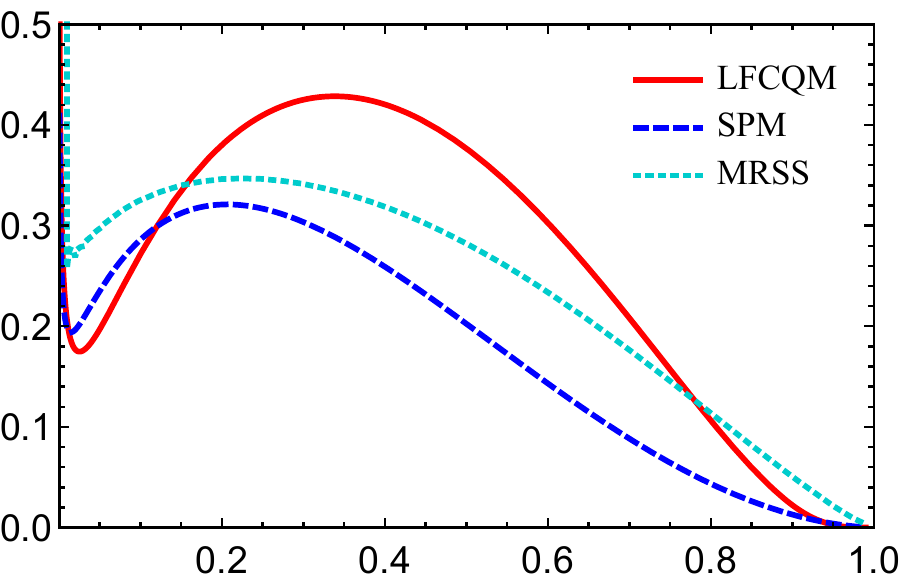}} & \hspace{5mm}\rotatebox[origin=c]{90}{$x \, h_{1,\, \pi^-}^{\perp (1) \bar u}$} &\vspace{-.5mm} \hspace{-3mm}\raisebox{-.5\height}{\includegraphics[width=0.33\textwidth]{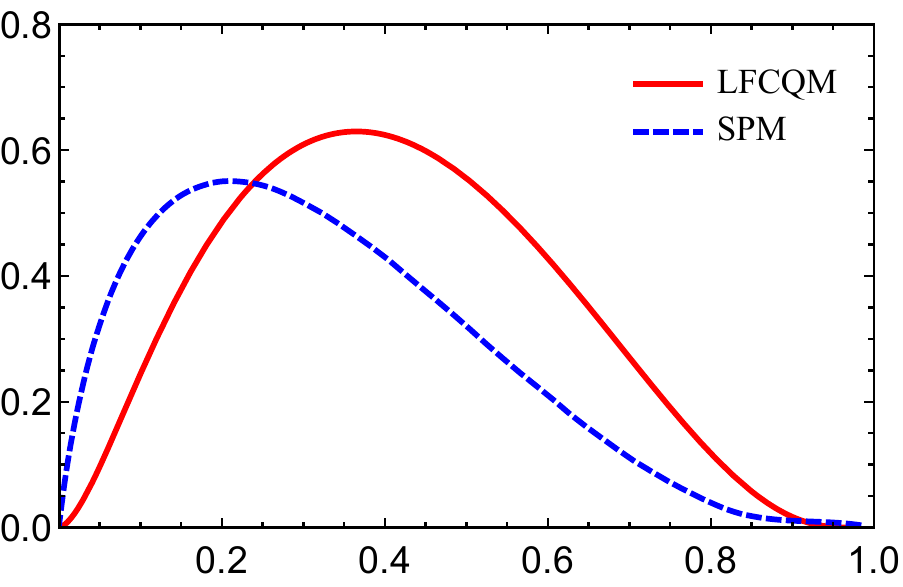}} \\
		& $x$ & & $x$ \\
		
	\end{tabular}
	\vspace{-3mm}
	\caption{\label{Fig:pion-models} 
        Left: $f^{\bar u}_{1,\pi^-}$ from LFCQM \cite{Pasquini:2014ppa} 
        and SPM~\cite{Gamberg:2009uk} LO-evolved to the scale $Q_0$ in comparison to MRSS parametrization
        \cite{Sutton:1991ay}. 
        Right: Predictions from LFCQM \cite{Pasquini:2014ppa} and
        SPM \cite{Gamberg:2009uk} for the pion Boer-Mulders function 
        (with the sign for DY) 
        for which no parametrizations are currently available.}
\end{figure}

\begin{figure}[t]
\centering
\vspace{-5mm}
	\begin{tabular}{ccccc}
	
	\hspace{-5mm}	\rotatebox[origin=c]{90}{$x \, f_{1,p}^u$} &\vspace{-.5mm} \hspace{-2mm}\raisebox{-.5\height}{\includegraphics[width=0.3\textwidth]{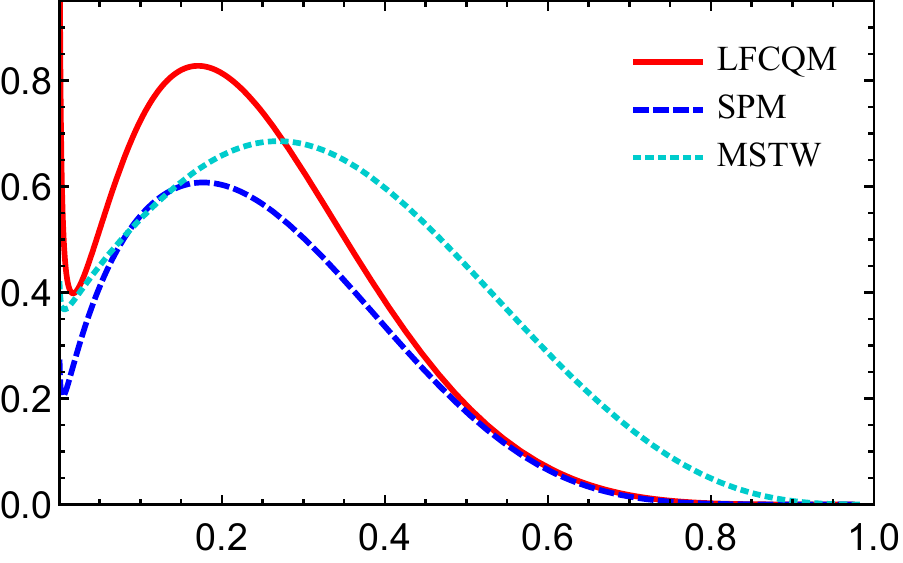}} & & \hspace{2mm} \rotatebox[origin=c]{90}{$x \, f_{1,p}^d$} &\vspace{-.5mm} \hspace{-2.5mm}\raisebox{-.5\height}{\includegraphics[width=0.3\textwidth]{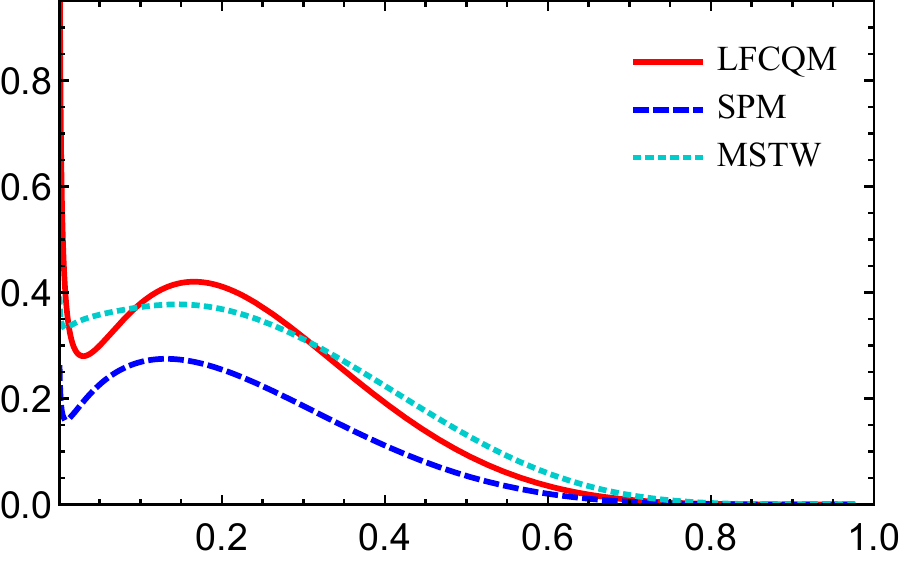}} \\
		& $x$ & & & $x$ \\	
		
	\hspace{-7mm}	\rotatebox[origin=c]{90}{$x \, f_{1T,p}^{\perp (1) u}$} &\vspace{-.5mm} \hspace{-3mm}\raisebox{-.5\height}{\includegraphics[width=0.308\textwidth]{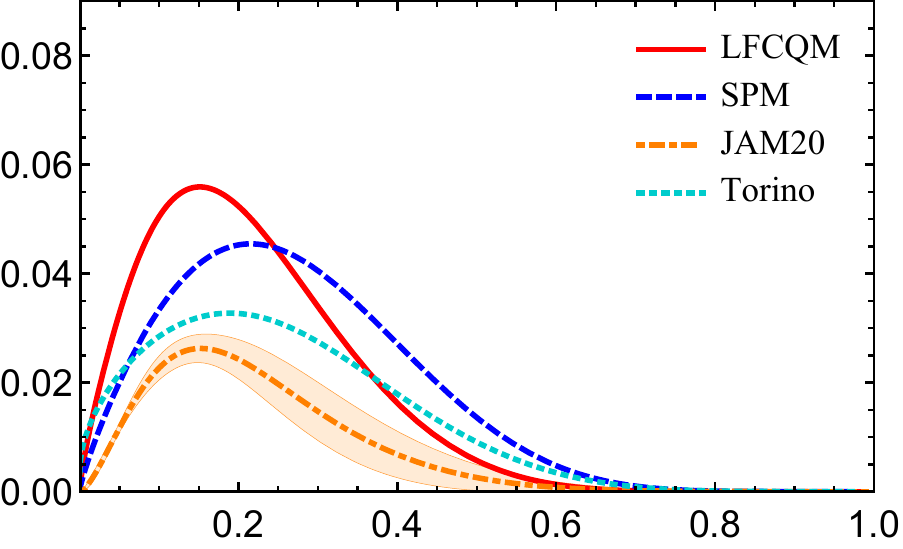}}  & & \rotatebox[origin=c]{90}{$x \, f_{1T,p}^{\perp (1) d}$} &\vspace{-.5mm} \hspace{-5mm}\raisebox{-.5\height}{\includegraphics[width=0.316\textwidth]{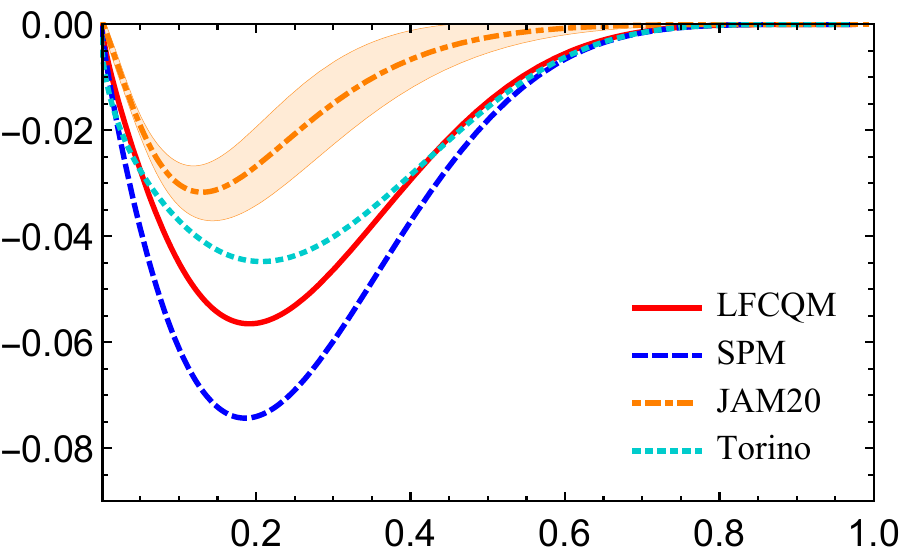}} \\
		& $x$ & & & $x$ \\		
		
	\hspace{-5mm} \rotatebox[origin=c]{90}{$x \, h_{1,p}^u$} &\vspace{-.5mm} \hspace{-2mm}\raisebox{-.5\height}{\includegraphics[width=0.3\textwidth]{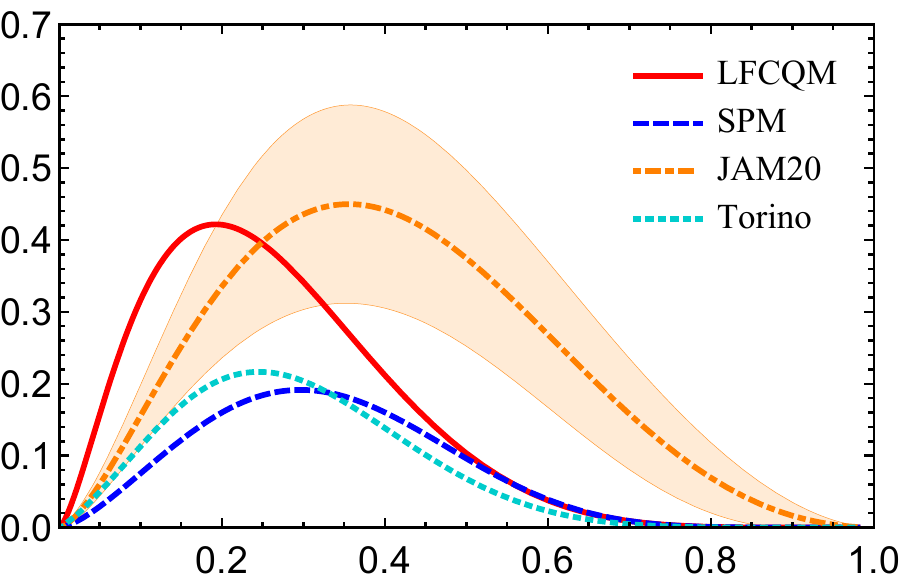}}  & & \rotatebox[origin=c]{90}{$x \, h_{1,p}^d$} &\vspace{-.5mm} \hspace{-4mm}\raisebox{-.53\height}{\includegraphics[width=0.309\textwidth]{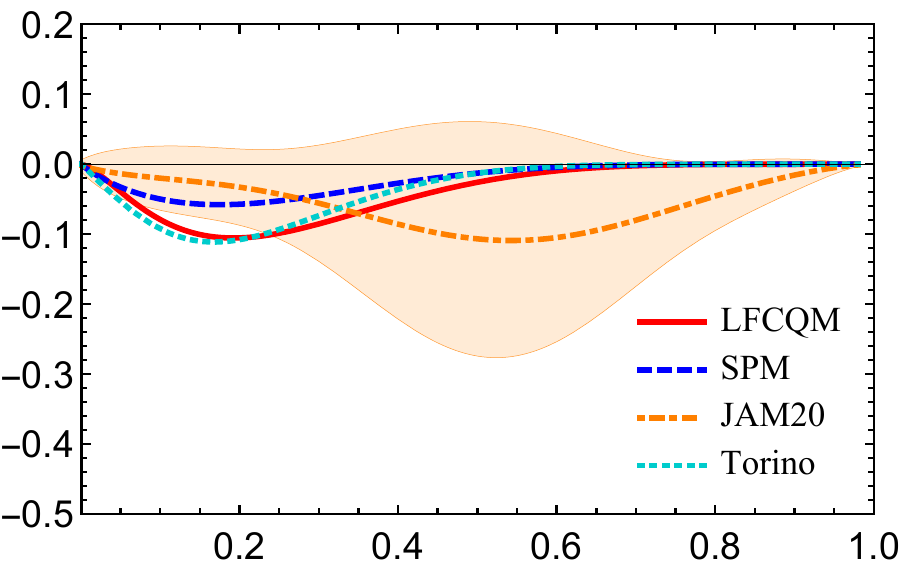}} \\
		& $x$ & & & $x$ \\		
		
	 \hspace{-7mm} \rotatebox[origin=c]{90} {$x \, h_{1,p}^{\perp (1) u}$} &\vspace{-.5mm} \hspace{-3mm}\raisebox{-.5\height}{\includegraphics[width=0.31\textwidth]{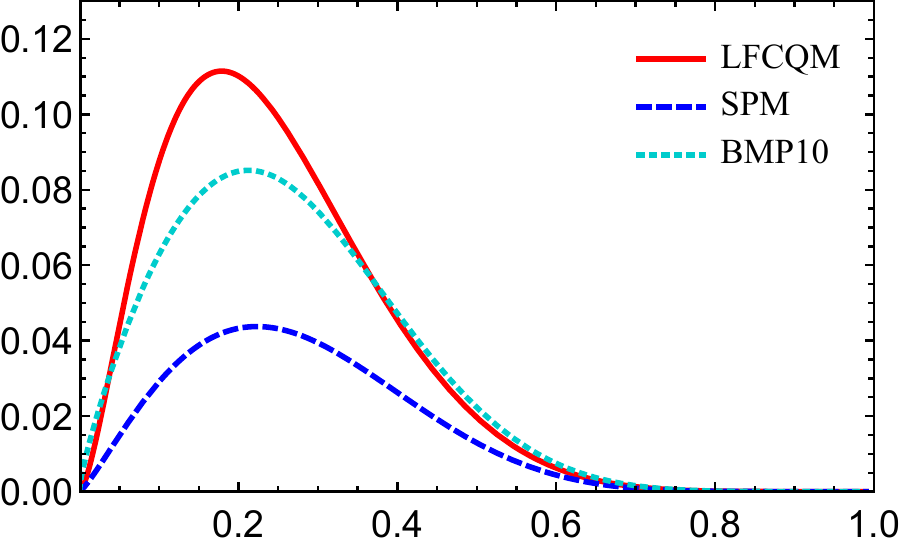}} & & \rotatebox[origin=c]{90}{$x \, h_{1,p}^{\perp (1) d}$} &\vspace{-.5mm} \hspace{-4mm}\raisebox{-.5\height}{\includegraphics[width=0.31\textwidth]{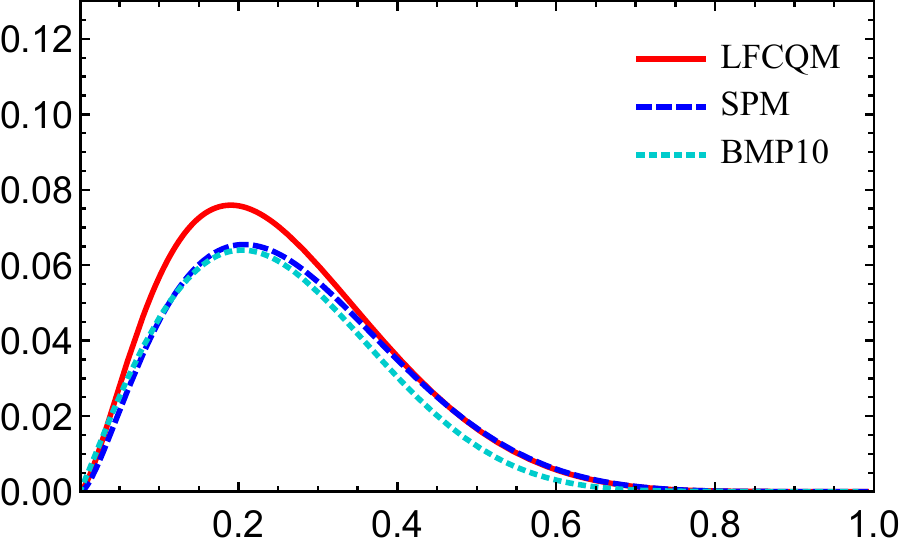}} \\
		& $x$ & & & $x$ \\		
		
	\hspace{-10mm}	\rotatebox[origin=c]{90}{$x \, h_{1T,p}^{\perp (2) u}$} &\vspace{-.5mm} \hspace{-6mm}\raisebox{-.5\height}{\includegraphics[width=0.326\textwidth]{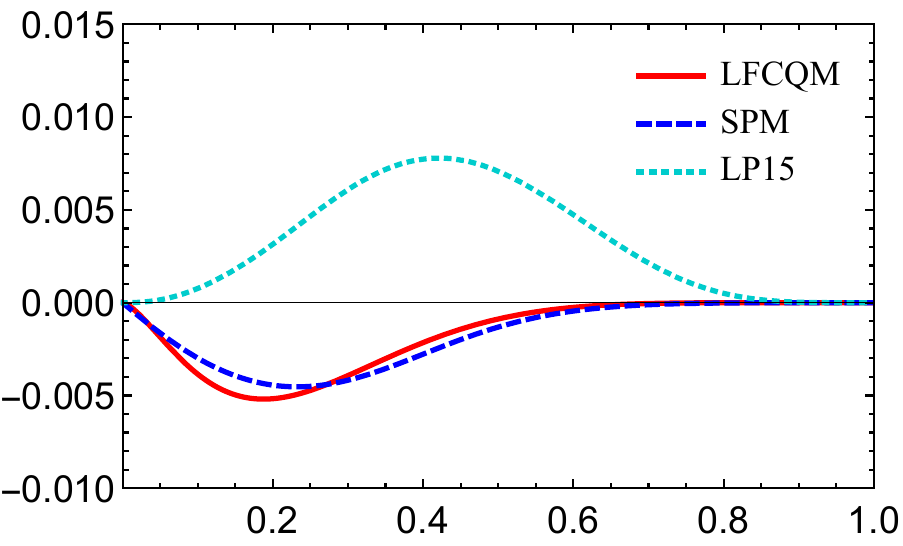}} & & \hspace{-1mm}\rotatebox[origin=c]{90}{$x \, h_{1T,p}^{\perp (2) d}$} &\vspace{-.5mm} \hspace{-6.5mm}\raisebox{-.5\height}{\includegraphics[width=0.326\textwidth]{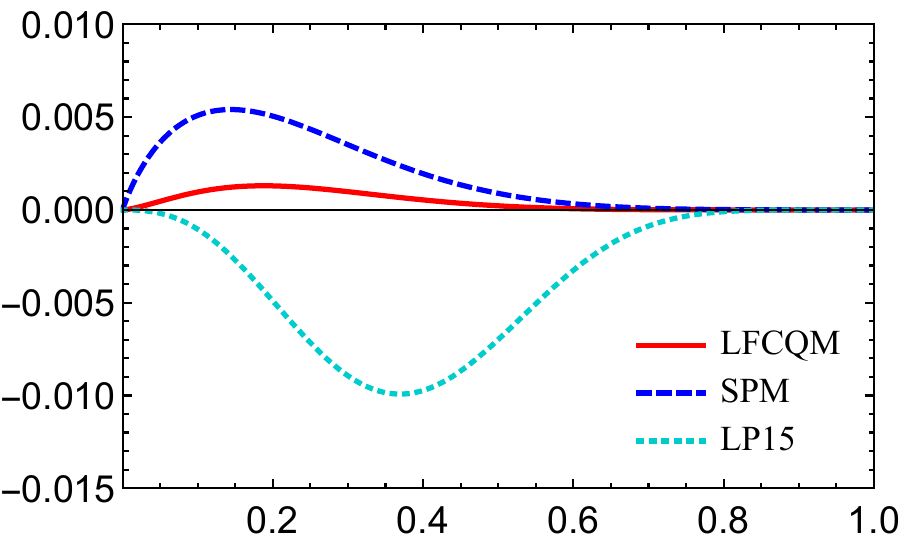}} \\
		& $x$ & & & $x$ \\		
		
	\hspace{-10mm} \rotatebox[origin=c]{90}{$x \, h_{1L,p}^{\perp (1) u}$} &\vspace{-.5mm} \hspace{-5mm}\raisebox{-.5\height}{\includegraphics[width=0.318\textwidth]{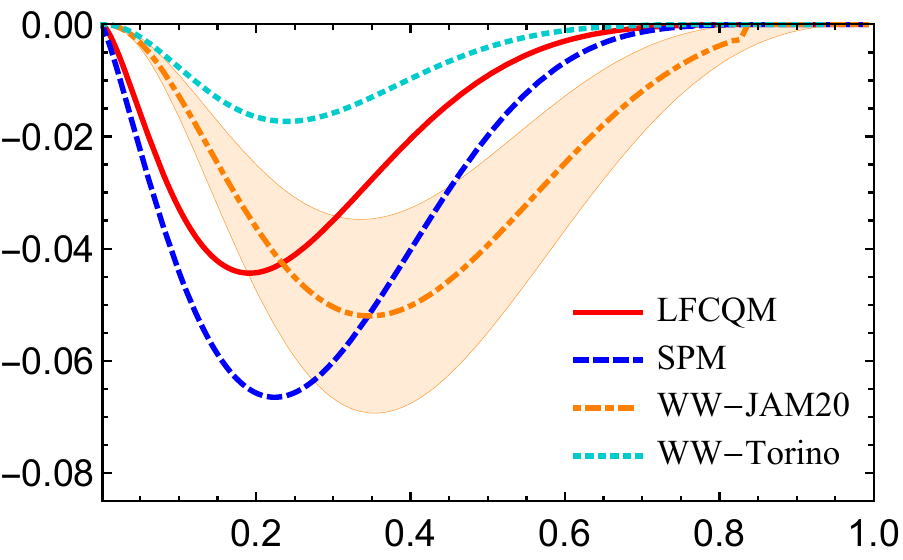}} & & \rotatebox[origin=c]{90}{$x \, h_{1L,p}^{\perp (1) d}$} &\vspace{-.5mm} \hspace{-5.5mm}\raisebox{-.5\height}{\includegraphics[width=0.318\textwidth]{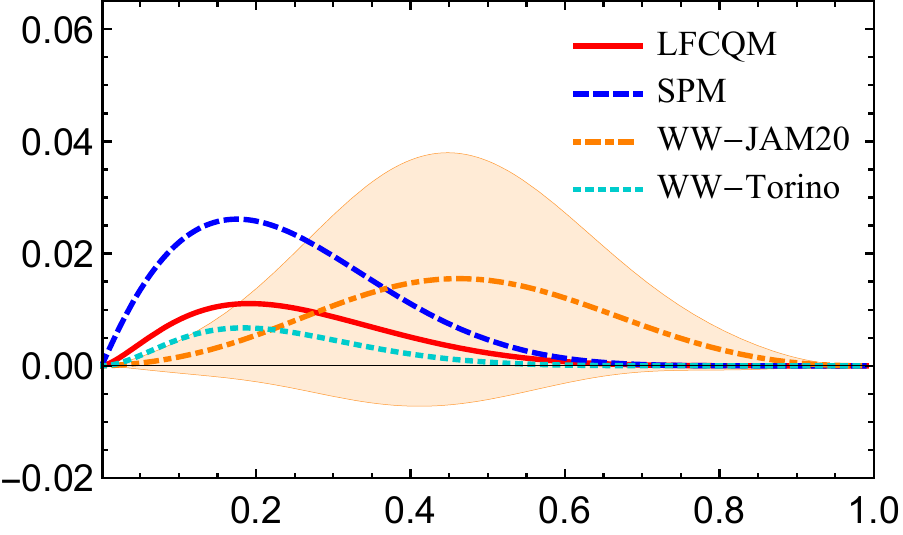}} \\
		& $x$ & & & $x$ \\		
				
	\end{tabular}
	
	\caption{\label{protonbasis-u+d} 
	The proton TMDs of $u$ and $d$ quarks in LFCQM \cite{Pasquini:2008ax,Boffi:2009sh,Pasquini:2011tk} 
	and SPM 
	\cite{Gamberg:2007wm} at 
	the scale $Q_0$ compared to phenomenological fits for $f_{1,p}$ from MSTW2008(LO) \cite{Martin:2009iq}, $f_{1T,p}^{\perp (1) a}$ from JAM20 \cite{Cammarota:2020qcw} and Torino \cite{Anselmino:2011gs}, $h_{1,p}^a$ from JAM20 \cite{Cammarota:2020qcw} and Torino \cite{Anselmino:2013vqa},
    $h_{1,p}^{\perp (1) a}$ from BMP10 \cite{Barone:2009hw}, $h_{1T,p}^{\perp (2) a}$ from LP15 \cite{Lefky:2014eia}. Sivers and Boer-Mulders TMDs are shown with the sign for DY process. The error bands show the 1-$\sigma$ uncertainty  of the JAM20 extractions \cite{Cammarota:2020qcw}.}   
    
\end{figure}

In this section we briefly review the two CQM frameworks, the LFCQM 
and the SPM, and compare them in Figs.~\ref{Fig:pion-models} and 
\ref{protonbasis-u+d} to the available phenomenological extractions 
used in this work.

Light-front models are based on the  decomposition of the hadron states in 
the  Fock space constructed in the framework of light-front quantization.
The hadron states are then obtained as a superposition of partonic
quantum states, each one multiplied by an  $N$-parton light-front wave function 
which gives the probability amplitude to find the corresponding $N$-parton
state in the hadron. In the LFCQM the light-front Fock expansion is truncated
to the leading component given by the valence $3q$ and $q\bar q$ contribution
in the proton and pion, respectively.
The  light-front wave functions
can be further decomposed in terms of light-front wave amplitudes
that are eigenstates of the total parton orbital angular momentum. 
The TMDs can then be expressed as overlap of light-front wave amplitudes 
with different orbital angular momentum~\cite{Pasquini:2008ax} which makes
very transparent the spin-orbit correlations encoded in the different TMDs~\cite{Pasquini:2008ax,Pasquini:2010af,Boffi:2009sh,Pasquini:2011tk,Pasquini:2014ppa}.
To model the $3q$ light-front wave function 
of the proton, the phenomenological Ansatz 
of Ref.~\cite{Schlumpf:1992pp} was used, describing the quark-momentum dependence
through a rational analytical expression with parameters fitted to the
anomalous magnetic moment of the proton and
neutron~\cite{Schlumpf:1992pp,Pasquini:2007iz}.
For the pion, the $q\bar q$ 
light-front wave function of Ref.~\cite{Schlumpf:1994bc} was used, 
with the quark-momentum dependent part given by a Gaussian function 
with parameters fitted to the pion charge radius and decay constant.

Spectator models are based on a field theoretical description of deep
inelastic scattering in a relativistic impulse approximation.
In this parton model-like factorization,  the cross section for deep
inelastic scattering processes  can be expressed in terms of a Born 
cross section and quark correlation functions~\cite{Gamberg:2005ip}.
In this framework, the quark correlation functions  are  hadronic matrix
elements expanded in Dirac and flavor structure multiplying form factors.
The essence of the SPMs  is to calculate the matrix elements of the quark
correlation function by the introduction of effective hadron-spectator-quark 
(e.g. nucleon-diquark-quark) vertices~\cite{Meyer:1990fr,Jakob:1997wg,Gamberg:2007wm} which in turn enable one to model  essential non-perturbative  
flavor and spin structure of hadrons.

The SPMs allow one to model the dynamics of  universality and
process dependence through studying the gauge-link, and phase content of  
TMDs~\cite{Brodsky:2002cx,Ji:2002aa,Goldstein:2002vv,Metz:2002iz,Gamberg:2003eg,Collins:2004nx,Gamberg:2008yt}.
In turn systematic phenomenological estimates for parton distributions
and fragmentation functions for both ``T-even'' and  ``T-odd'' TMDs 
have been  carried out~\cite{Ji:2002aa,Goldstein:2002vv,Gamberg:2003ey,Boer:2002ju,Gamberg:2007wm,Bacchetta:2008af}. 
In regard to the latter, it is in this framework that the first calculations 
of the Sivers and Boer-Mulders functions of the nucleon were carried
out~\cite{Brodsky:2002cx,Ji:2002aa,Goldstein:2002vv}
and shown on general grounds to contribute to semi-inclusive processes 
at leading power in the hard scale. Later the Boer-Mulders function 
of the pion was calculated in Ref.~\cite{Gamberg:2009uk}. 
The model parameters are determined by comparing the 
SPM results for $f_{1,p}^u(x)$ and $f_{1,p}^d(x)$ 
to the LO low-scale ($\mu^{2}_0=0.26$ GeV$^2$) 
GRV98 parametrization~\cite{Gluck:1998xa}.

The proton TMDs for $u$- and $d$- quarks are given by linear
combinations of contributions from axial-vector and scalar 
diquarks assuming $\rm SU(2)$ flavor symmetry
\cite{Jakob:1997wg,Gamberg:2007wm}.

We choose the scale
$Q_0^2 = 2.4\,{\rm GeV}^2$
as the initial scale for the CSS evolution.
The evolution effects between the initial model scale $\mu_0\sim 0.5\,{\rm GeV}$ and $Q_0$ cannot be determined 
exactly in the CSS formalism, see section~\ref{sec-2.3}, and they also cannot be neglected. 
We therefore estimate them as follows.  
We start with the model predictions for the parton
distributions or transverse moments of TMDs as they 
appear in Eq.~(\ref{Eq:Gauss}) at the initial quark model 
scale $\mu_0$. We evolve them using LO DGLAP evolution to 
the scale $Q_0$. In contrast to CSS, experience with 
implementing DGLAP evolution at low scales is available \cite{Gluck:1991ng,Gluck:1991ey,Gluck:1994uf,Gluck:1998xa,Gluck:1999xe}. 
Hereby we use exact DGLAP evolution for $f_{1,h}^a(x)$ and $h_{1,p}^a(x)$.
In all other cases we use approximate DGLAP evolution: for the transverse
momenta of the proton Sivers function we use the $f_{1,h}^a(x)$-nonsinglet 
evolution shown to lead good results in the LFCQM model study of SIDIS 
asymmetries \cite{Pasquini:2011tk}, while for all the chiral-odd TMDs we assume 
the DGLAP evolution of transversity \cite{Hirai:1997mm,Boffi:2009sh}.
For the $k_T$-dependencies of the TMDs we use the same
input from TMD parametrizations as described in Eq.~(\ref{Eq:Gauss}).

The predictions from both models evolved in this way 
are shown along with the available parametrizations in
Figs.~\ref{Fig:pion-models}-\ref{protonbasis-u+d} at the scale $Q_0$. 
It is important to stress that in this way we are able to test the
$x$-dependencies of the model predictions against the COMPASS data. 
The ultimate goal would be to test similarly also the quark model 
predictions for $k_T$-dependencies. This requires an implementation of 
the CSS evolution starting from low initial scales $\mu_0< 1\,{\rm GeV}$
which is beyond the scope of this work, and will be addressed in future studies.

The result from the LFCQM~\cite{Pasquini:2014ppa} and the SPM~\cite{Gamberg:2009uk} for $f_{1,\pi^-}^{\bar u}(x)$ (which coincides with $f_{1,\pi^-}^{d}(x)$ due to isospin symmetry) compare well 
to the MRSS parametrization \cite{Martin:2009iq},
see Fig.~\ref{Fig:pion-models}.
In the region $0.2\lesssim x_\pi \lesssim 0.6$,
in which the COMPASS Drell-Yan data points lie, 
the SPM result agrees within 20-40$\,\%$ with MRSS \cite{Martin:2009iq}.
The two models agree well with each other 
in the case of the pion Boer-Mulders TMD 
$h_{1,\pi^-}^{\perp(1) \bar u}(x)=h_{1,\pi^-}^{\perp(1) d}(x)$
for which no extraction is available (so far). 
This robustness of the model predictions is important: the pion Boer-Mulders
function enters 4 (out of 6) twist-2 pion-nucleon DY structure functions.

The results from the LFCQM~\cite{Pasquini:2008ax,Boffi:2009sh,Pasquini:2011tk}  and the SPM~\cite{Gamberg:2007wm} for the proton quark distributions are 
shown in Fig.~\ref{protonbasis-u+d}. 
The region $0.05\lesssim x_p\lesssim0.4$
is probed in the COMPASS DY measurements~\cite{Aghasyan:2017jop}, see section~\ref{section:kinematics}. The model results for the functions
$f_{1,p}^u(x)$, $f_{1,p}^d(x)$, 
$f_{1T,p}^{(1)u}(x)$, $f_{1T,p}^{(1)d}(x)$, 
$h_{1,p}^d(x)$, $h_{1,p}^{\perp(1)d}(x)$, 
$h_{1L,p}^{(1)u}(x)$, $h_{1T,p}^{(2)u}(x)$
agree within 20-40$\,\%$, and for 
$h_{1,p}^u(x)$, 
$h_{1,p}^{\perp(1)u}(x)$, 
$h_{1L,p}^{(1)d}(x)$
within 40-60$\,\%$.
Merely for $h_{1T,p}^{(2)d}(x)$ we observe a more sizable 
spread of model predictions. 
In all cases the models agree on the signs of the TMDs.
The model results for the unpolarized distributions agree 
reasonably well with MSTW \cite{Martin:2009iq}. 
The model predictions for transversity and Sivers function are compatible 
with the corresponding Torino  \cite{Anselmino:2011gs,Anselmino:2013vqa} 
and JAM20 fits \cite{Cammarota:2020qcw}.
The 1-$\sigma$ uncertainty bands are shown for JAM20 \cite{Cammarota:2020qcw}.
The corresponding uncertainty bands of the Torino parametrizations
\cite{Anselmino:2011gs,Anselmino:2013vqa} are somewhat larger 
(as more data were used in the JAM20 analysis, cf.\ section~\ref{sec-2.3})
and not displayed for better visibility. The proton Boer-Mulders 
function from models is in good agreement with the BMP10
extraction~\cite{Barone:2009hw} which has significant statistical and
systematic uncertainties, as discussed in section~\ref{sec-2.3}, and 
are not shown in Fig.~\ref{protonbasis-u+d}. 
The model predictions for pretzelosity show little agreement with the 
best fit result from LP15 \cite{Lefky:2014eia}, but are within its
1-$\sigma$ region which is not shown in the plot.

The comparison in Fig.~\ref{protonbasis-u+d} indicates an accuracy 
of the CQMs which is in many cases of the order of  20--40$\,\%$.
Considering the much different physical foundations of the two models,
one may speak about an overall robust CQM picture for the TMDs
needed in our work.

\section{Results and observations}
\label{sec-3}

In this section we briefly describe the COMPASS experiment,
outline how we explore the model predictions and phenomenological 
TMD fits, present our results, and compare them to the data.

\subsection{The COMPASS Drell-Yan experiment}
\label{section:kinematics}

The COMPASS 2015 data \cite{Aghasyan:2017jop} were taken with 
a pion beam  of $190 \; {\rm GeV}$ impinging on a transversely 
polarized NH$_3$ target with a polarization of $\la S_T\ra \approx 73\,\%$ 
and a dilution factor $\la f\ra \approx 0.18$.
The dimuon mass range $4.3 \; {\rm GeV}< Q <8.5 \; {\rm GeV}$ above  
charmonium resonance region but below $\Upsilon$ threshold was covered 
with the mean value $\la Q\ra=5.3\,{\rm GeV}$.  
Due to the fixed target kinematics the pion structure was probed at 
higher $\la x_\pi\ra =0.50$ compared to the proton $\la x_p\ra = 0.17$.
The cut $q_T>0.4\,{\rm GeV}$ was imposed and $\la q_T\ra = 1.2\,{\rm GeV}$ \cite{Aghasyan:2017jop}. The analysis of the data collected by the 
experiment in 2018 in similar conditions is currently under way~\cite{Parsamyan:2019wwd}.

\subsection{The approaches for numerical estimates}
\label{Sec3.2-approaches}

The Sivers asymmetry $A_{UT}^{\sin\phi}$ can be described completely
in terms of both, model predictions and available parametrizations, and is
the only asymmetry where the latter is possible. For the phenomenological
calculation we will use the Torino \cite{Anselmino:2013vqa} and JAM20
\cite{Cammarota:2020qcw} analysis results for $f_{1T.p}^{\perp(1)a}(x)$, 
and MSTW \cite{Martin:2009iq} and MRSS \cite{Sutton:1991ay} 
parametrizations for proton and pion collinear unpolarized distributions. 

The other asymmetries require the knowledge of the pion Boer-Mulders
function for which no parametrization is available. In these cases
we shall adopt two different main approaches, pure and hybrid,
see Fig.~\ref{Fig-02:knowledge-TMDs} for an overview.
We will present therefore up to 
four different calculations for each observable by exploring 
the model results and available parametrizations discussed in sections~\ref{sec-2.2},~\ref{sec-2.3},~\ref{sec-2.4} and displayed in Figs.~\ref{Fig:pion-models}-\ref{protonbasis-u+d}.
The first approach makes a pure use of model predictions for all 
pion and proton TMDs which will be labelled in the plots by the
acronyms LFCQM or SPM.

In the hybrid-approaches we will use the minimal model input,
the predictions from the LFCQM~\cite{Pasquini:2014ppa} and SPM~\cite{Gamberg:2009uk}
for the pion Boer-Mulders
function, and the maximal input from parametrizations: 
JAM20 \cite{Cammarota:2020qcw} for $f_{1T,p}^{\perp a}$ and $h_{1,p}^a$,
BMP10 \cite{Barone:2009hw} for $h_{1,p}^{\perp a}$, and 
LP15 \cite{Lefky:2014eia} for $h_{1T,p}^{\perp a}$. 
The results will be labelled respectively
as ``LFC-JAM20'', ``LFC-LP15'', ``LFC-BMP10''  
or ``SPM-JAM20,'' ``SPM-LP15,'' ``SPM-BMP10.''
For $h_{1L,p}^{\perp a}$ we make use of WW-type approximation 
which allows one to approximate this TMD in terms of $h_{1,p}^a$ for
which we will use JAM20 \cite{Cammarota:2020qcw}. 
WW-type approximations  were explored in Ref.~\cite{Bastami:2018xqd} 
and shown to work well with the available data. We will add ``WW'' 
in the label of calculation when WW approximation is used.
For all hybrid calculations we will use the parametrizations
\cite{Martin:2009iq,Sutton:1991ay} for $f_{1,p}^a$ and $f_{1,\pi}^a$.

\begin{figure}[t]
\centering
  \begin{tabular}{cccc}
		 \rotatebox[origin=c]{90}{{$A_{UT}^{\sin \phi_S}$}} &  \hspace{-2.5mm}\raisebox{-.5\height}{\includegraphics[width=4.5cm]{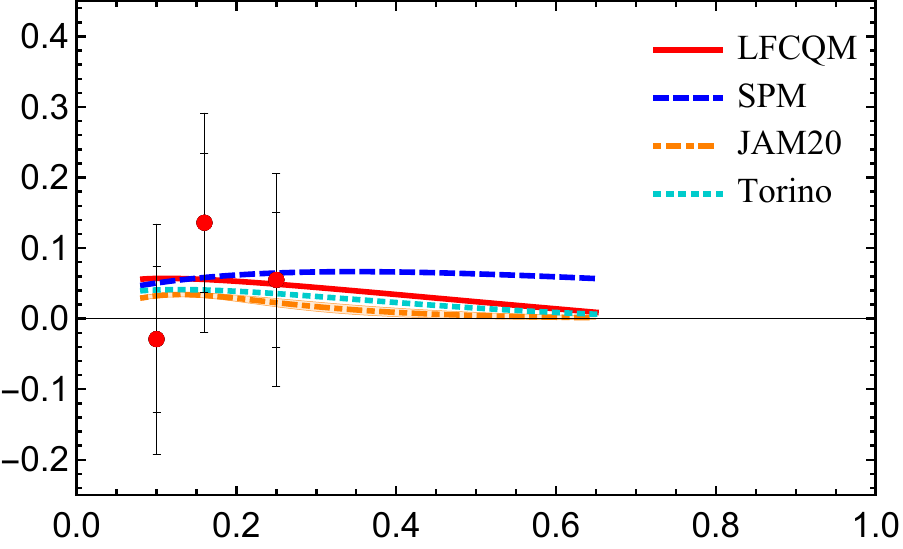}} &  \hspace{-2.5mm}\raisebox{-.5\height}{\includegraphics[width=4.5cm]{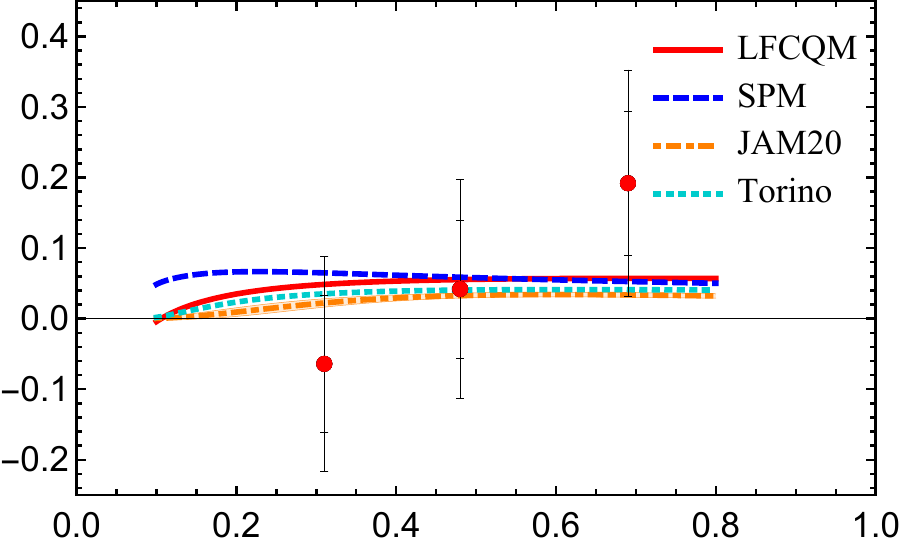}} & \hspace{-2.5mm}\raisebox{-.5\height}{\includegraphics[width=4.5cm]{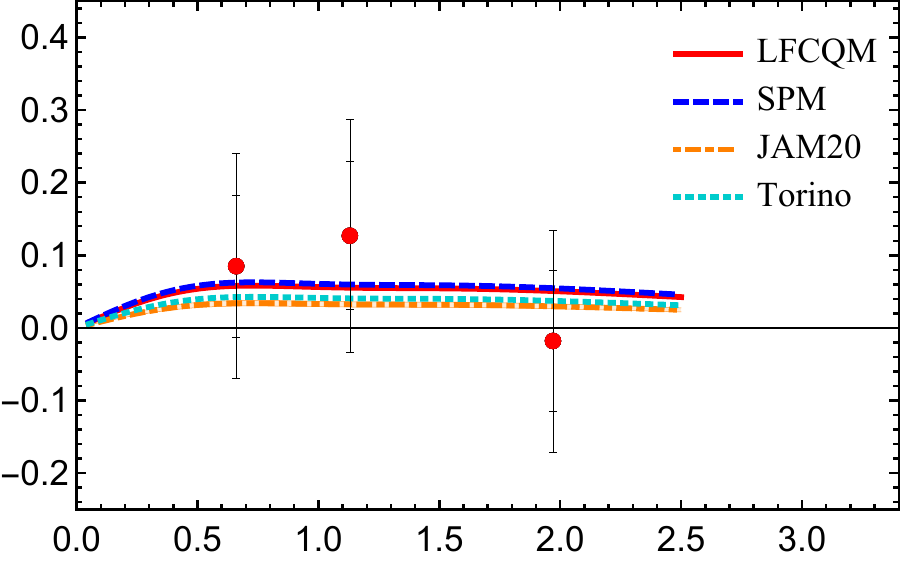}} \\
		& $x_p$ & $x_\pi$ & $q_T$ \\	
  \end{tabular}
  
 \vspace{-3mm}
\caption{\label{aut1x-COM} 
  $A_{UT}^{\sin\phi}$ as a function of 
$x_p$ (left), $x_{\pi}$ (middle) and $q_T$ (right)
  vs COMPASS data \cite{Aghasyan:2017jop}.}
  
\vspace{7mm}  

\centering
  \begin{tabular}{cccc}
		 \rotatebox[origin=c]{90}{{$A_{UT}^{\sin (2\phi-\phi_S)}$}} &  \hspace{-2.5mm}\raisebox{-.5\height}{\includegraphics[width=4.5cm]{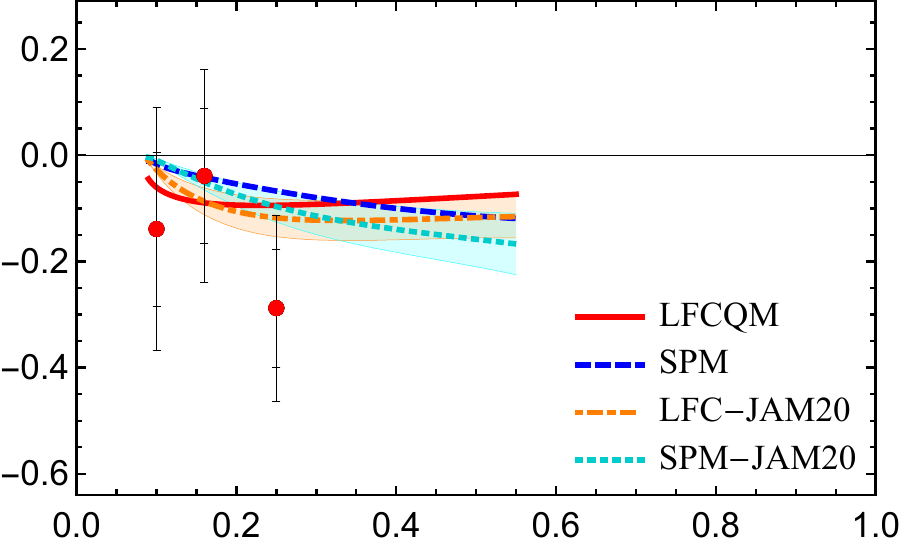}} &  \hspace{-2.5mm}\raisebox{-.5\height}{\includegraphics[width=4.5cm]{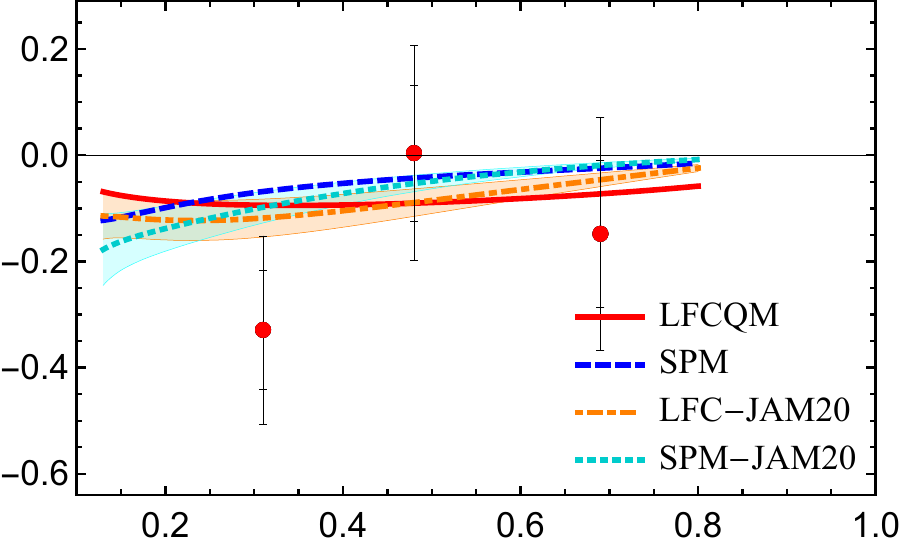}} &  \hspace{-2.5mm}\raisebox{-.5\height}{\includegraphics[width=4.5cm]{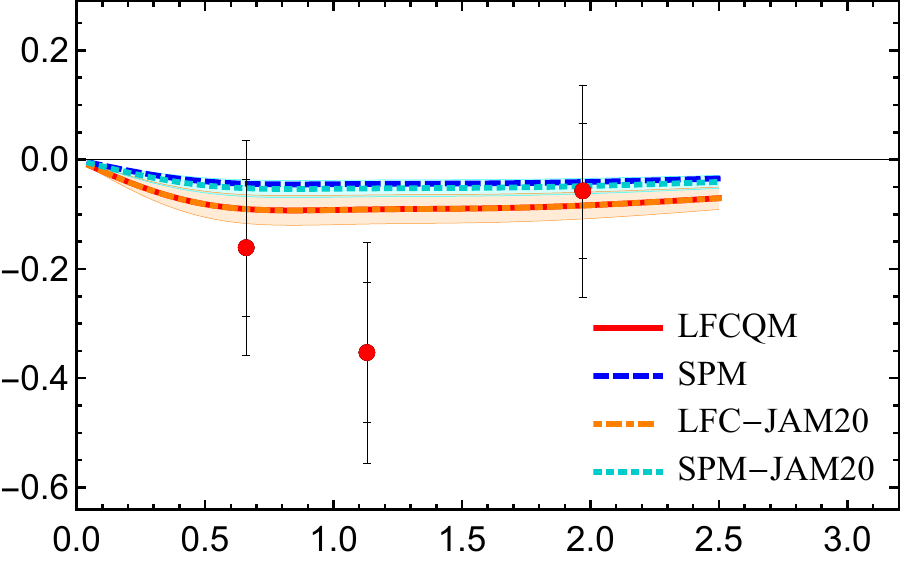}} \\
		& $x_p$ & $x_\pi$ & $q_T$ \\	 
  \end{tabular}
\vspace{-3.5mm}
\caption{\label{autsin2pmpx-COM} 
  $A_{UT}^{\sin(2\phi-\phi_S)}$ as a function of 
$x_p$ (left), $x_{\pi}$ (middle) and $q_T$ (right)
  vs COMPASS data \cite{Aghasyan:2017jop}.}

\vspace{7mm}

  \begin{tabular}{cccc}
		\rotatebox[origin=c]{90}{{$A_{UT}^{\sin (2\phi+\phi_S)}$}} &  \hspace{-2.5mm}\raisebox{-.5\height}{\includegraphics[width=4.5cm]{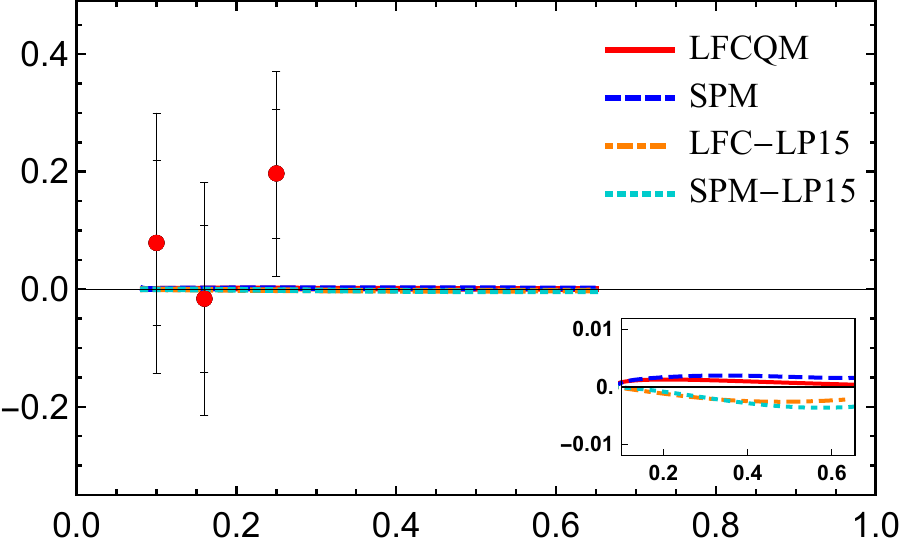}} &  \hspace{-2.5mm}\raisebox{-.5\height}{\includegraphics[width=4.5cm]{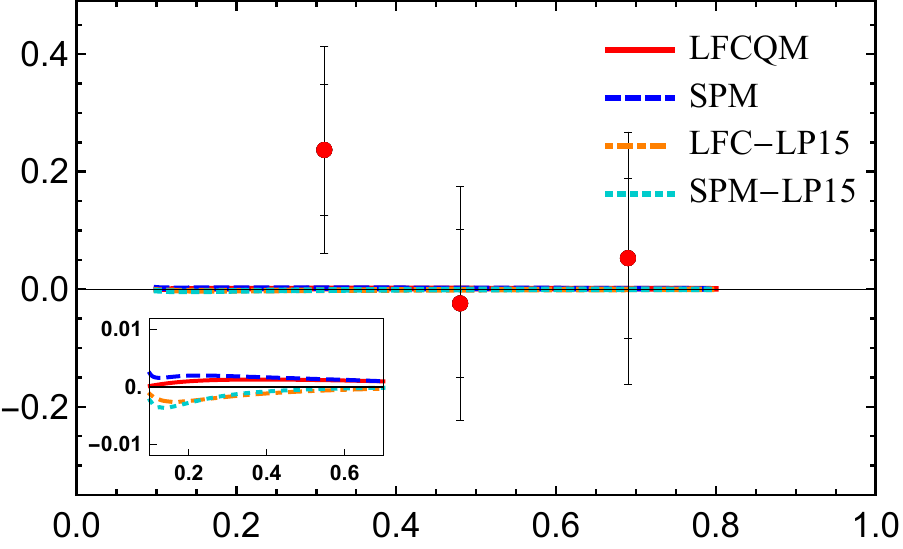}} & \hspace{-2.5mm}\raisebox{-.5\height}{\includegraphics[width=4.5cm]{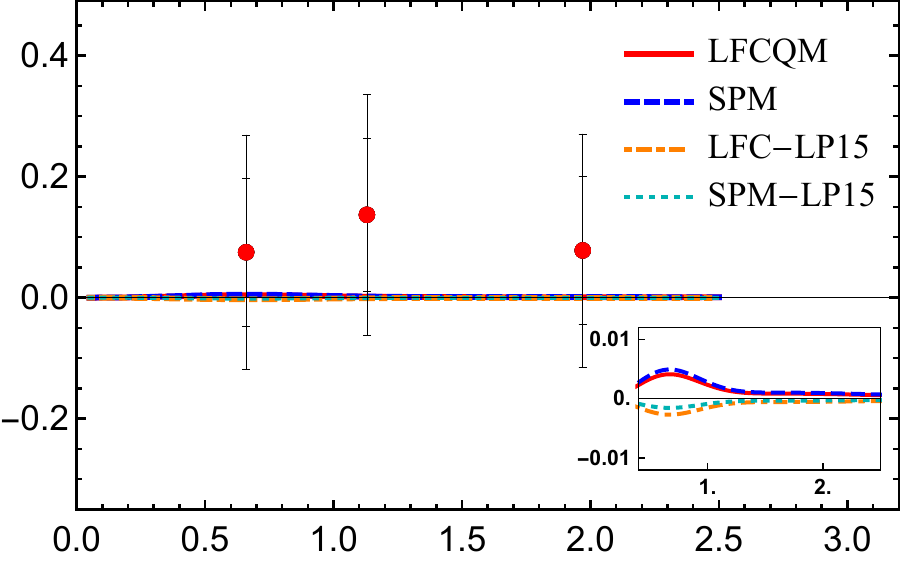}} \\
		& $x_p$ & $x_\pi$ & $q_T$ \\	 
  \end{tabular}
\vspace{-3.5mm}
\caption{\label{autsin2pppx-COM} 
  $A_{UT}^{\sin(2\phi+\phi_S)}$ as a function of 
$x_p$ (left), $x_{\pi}$ (middle) and $q_T$ (right)
  vs COMPASS data \cite{Aghasyan:2017jop}.}

\vspace{7mm}

  \begin{tabular}{cccc}
		 \rotatebox[origin=c]{90}{{$A_{UU}^{\cos 2\phi}$}} &  \hspace{-2.5mm}\raisebox{-.5\height}{\includegraphics[width=4.5cm]{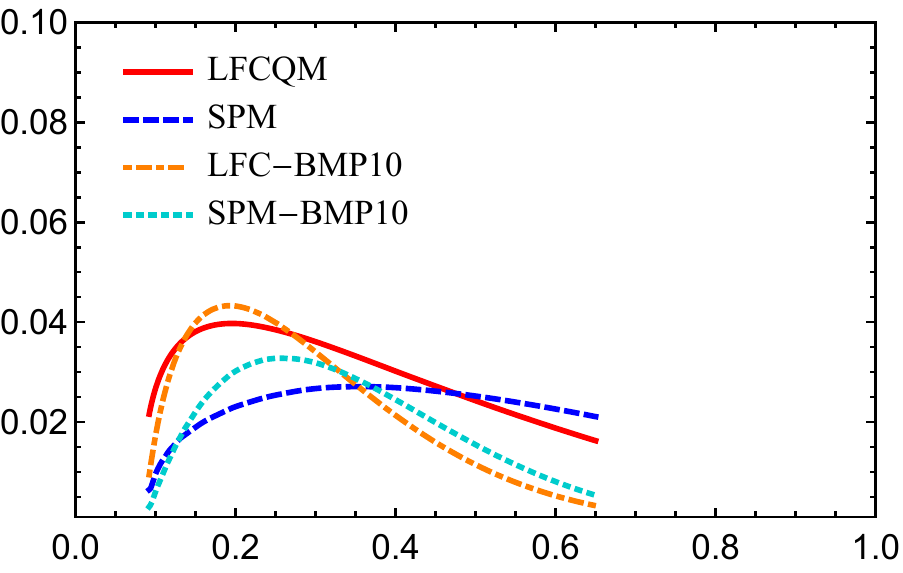}} &  \hspace{-2.5mm}\raisebox{-.5\height}{\includegraphics[width=4.5cm]{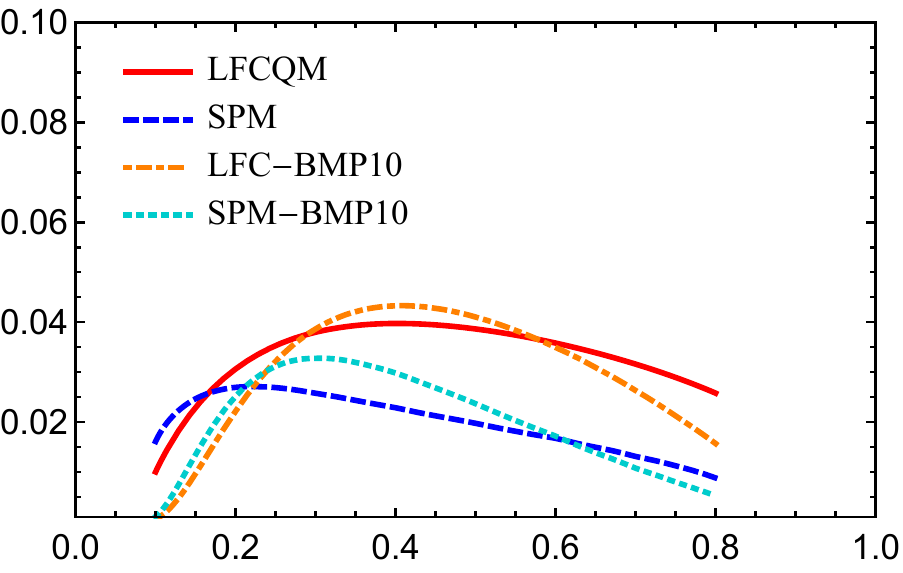}} &  \hspace{-2.5mm}\raisebox{-.5\height}{\includegraphics[width=4.5cm]{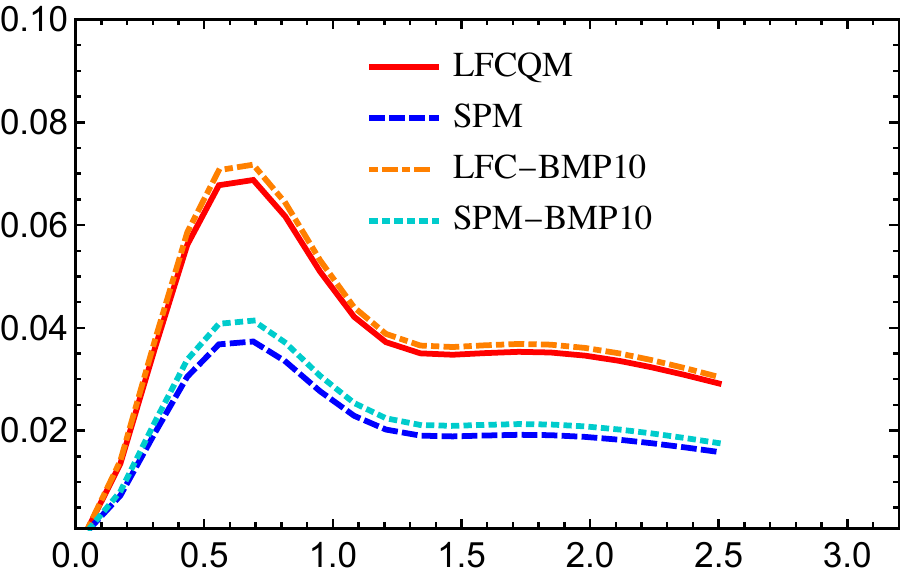}} \\
		& $x_p$ & $x_\pi$ & $q_T$ \\	
  \end{tabular}
\vspace{-3.5mm}
\caption{\label{auucos2px-COM} 
  $A_{UU}^{\cos 2\phi}$ as a function of 
$x_p$ (left), $x_{\pi}$ (middle) and $q_T$ (right)
  in the COMPASS kinematics.}
  
\vspace{7mm}

  \begin{tabular}{cccc}		
		 \rotatebox[origin=c]{90}{{$A_{UL}^{\sin 2\phi}$}} & \hspace{-2.5mm}\raisebox{-.5\height}{\includegraphics[width=4.5cm]{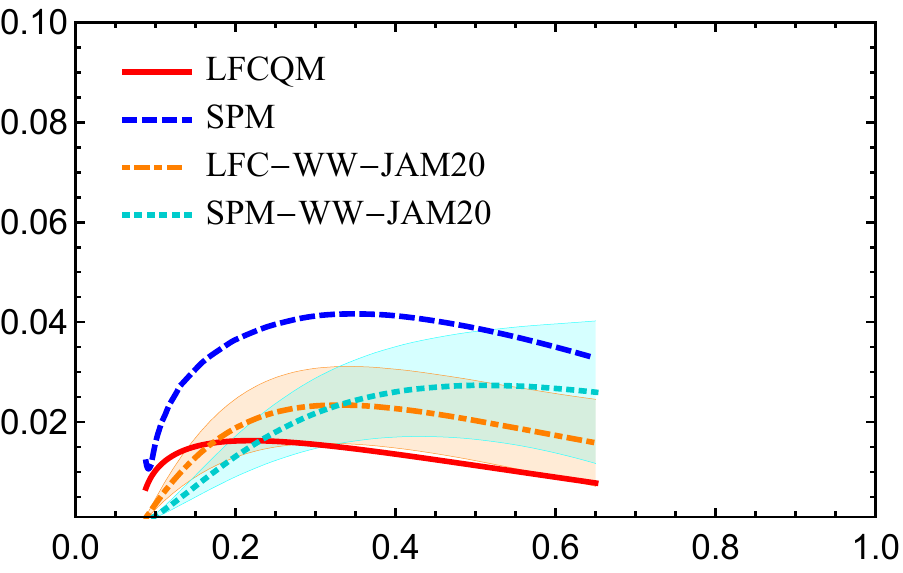}} &  \hspace{-2.5mm}\raisebox{-.5\height}{\includegraphics[width=4.5cm]{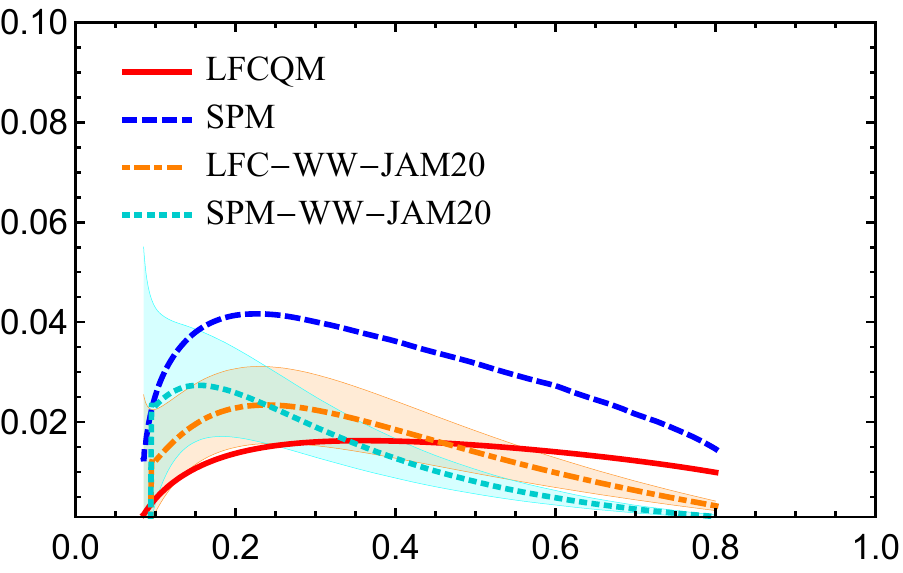}} &  \hspace{-2.5mm}\raisebox{-.5\height}{\includegraphics[width=4.5cm]{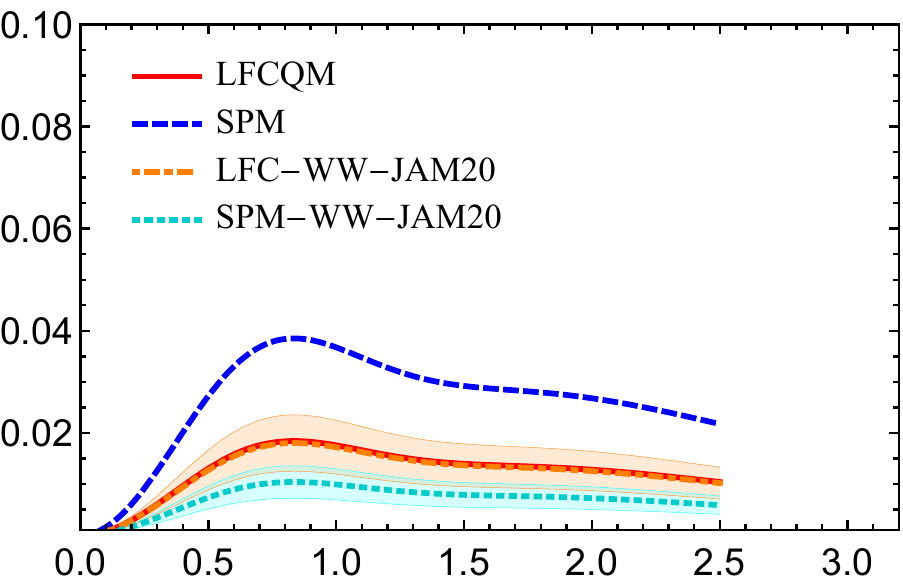}} \\
		& $x_p$ & $x_\pi$ & $q_T$ \\	
  \end{tabular}
  
  \vspace{-3.5mm}
\caption{\label{aulsin2px-COM} 
  $A_{UL}^{\sin 2\phi}$ as a function of 
$x_p$ (left), $x_{\pi}$ (middle) and $q_T$ (right)
  in the COMPASS kinematics. }
\end{figure}

\subsection{Discussion of the results and comparison to available data}
\label{sec-3.2}

Numerical results for the leading-twist pion-nucleon DY 
asymmetries are shown in Figs.~\ref{aut1x-COM}--\ref{aulsin2px-COM}
in comparison to available COMPASS data. 
Table~\ref{table-overview} gives a detailed overview 
on the model results and phenomenological information. 

Let us start the discussion with the Sivers asymmetry. One of the most striking features of 
 ``naively'' T-odd (Sivers, Boer-Mulders) TMDs is the expected sign 
change~\cite{Collins:2002kn} from SIDIS to DY due to the 
difference of initial  (DY) versus final (SIDIS) state interactions
\cite{Brodsky:2002rv,Boer:2002ju}.
Verification of the sign change of the Sivers function is one of the milestones of DY programs of COMPASS and RHIC~\cite{Aschenauer:2015eha}.
In SIDIS the proton $u$-quark Sivers function is negative,
while STAR RHIC \cite{Adamczyk:2015gyk} $W^\pm/Z$ 
asymmetry data favor a positive sign~\cite{Anselmino:2016uie} hinting on the predicted 
process dependence of T-odd TMDs \cite{Collins:2002kn}.

The predictions for the $A_{UT}^{\sin\phi_S}$ asymmetry at COMPASS are positive, 
see for instance 
Refs.~\cite{Efremov:2004tp,Collins:2005rq,Anselmino:2009st}.
Our calculations confirm this expectation, see 
Fig.~\ref{aut1x-COM} where we compare our results 
to COMPASS data~\cite{Aghasyan:2017jop}.
The $u$-quark Sivers function in DY is expected to be positive, 
see Fig.~\ref{protonbasis-u+d}. 
If we disregard sea quark effects, which were shown to play a 
negligible role in $\pi^-$-proton DY in the COMPASS kinematics
\cite{Collins:2005rq}, then 
$A_{UT}^{\sin\phi_S} \propto f_{1T,p}^{\perp u} (x_p) > 0$.
The experimental error bars are currently sizeable, but the data 
show a tendency to positive asymmetry, see  Fig.~\ref{aut1x-COM}, 
in agreement with the expected sign change of the Sivers function.
Clearly, more experimental evidence is needed to corroborate 
this finding.

In the global QCD analysis of single-spin asymmetries
\cite{Cammarota:2020qcw} the COMPASS data~\cite{Aghasyan:2017jop} 
were used, such that the  JAM20 result in Fig.~\ref{aut1x-COM}
is consistent with {\em all} present-day data on  observables related
to Sivers functions. It is worth remarking that predictions based 
on the earlier Torino extraction \cite{Anselmino:2011gs}
(which used SIDIS data only) yield a somewhat larger 
asymmetry than JAM20 and are closer to the LFCQM and SPM
results in Fig.~\ref{aut1x-COM}.
This result is consistent with the different size of Sivers functions found in Ref.~\cite{Anselmino:2011gs} and Ref.~\cite{Cammarota:2020qcw}, see Fig.~\ref{protonbasis-u+d}.

Fig.~\ref{autsin2pmpx-COM} shows the asymmetry 
$A_{UT}^{\sin(2\phi-\phi_S)}$ which arises from a convolution
of transversity and pion Boer-Mulders function in comparison
to COMPASS data \cite{Aghasyan:2017jop}. 
In the case of this asymmetry the pure model and hybrid 
calculations yield results in good mutual agreement. 
Neglecting sea quarks, it is
$A_{UT}^{\sin(2\phi-\phi_S)}\propto - h_{1, \pi^-}^{\perp (1)\bar u }(x_\pi) h_{1,p}^u (x_p)<0$. 
Both, $h_{1, \pi^-}^{\perp (1)\bar u }$ and $h_{1,p}^u$ are positive, see Fig~\ref{protonbasis-u+d}, and we predict a 
negative asymmetry. This is consistent with the trend of the data.
We therefore conclude that the COMPASS data~\cite{Aghasyan:2017jop}
indicate a positive sign for the pion Boer-Mulders 
TMD~$h_{1, \pi^-}^{\perp (1)\bar u }$.
(It~is~important to recall that absolute signs in extractions
of chiral-odd TMDs and fragmentation functions are convention-dependent
because chiral-odd functions contribute to observables always in
connection with other chiral-odd functions. The convention used 
for TMD extractions is $h_{1,p}^u(x)>0$. This sign is a choice 
which is well-informed by model and lattice QCD calculations
but not an experimental observation.)
The indication that $h_{1, \pi^-}^{\perp (1)\bar u }>0$ is
an important result which can be used to test the process dependence
of the proton Boer-Mulders function, see below.

Fig.~\ref{autsin2pppx-COM} shows  
$A_{UT}^{\sin(2\phi+\phi_S)}$ which is due to the convolution
of pretzelosity and pion Boer-Mulders function compared to  COMPASS
data \cite{Aghasyan:2017jop}. This asymmetry is proportional to
$q_T^3$ for $q_T\ll \,{1\,\rm GeV}$. This leads to a kinematic
suppression of this asymmetry as compared to the two previous 
asymmetries (both proportional to $q_T$ at small transverse momenta). 
As a consequence $A_{UT}^{\sin(2\phi+\phi_S)}$ is by far the 
smallest of the leading-twist asymmetries in pion-nucleon DY. 
Numerically it is $1\,\%$ or smaller, such that we had to include the
insets in Fig.~\ref{autsin2pppx-COM} to display the theoretical curves.
The LFCQM and the SPM are in good agreement with each other, 
but not with the LP15 fit of pretzelosity \cite{Lefky:2014eia} which
suggests an opposite sign for the asymmetry.
At this point one has
to stress that the LP15 fit of \cite{Lefky:2014eia} has a large 
statistical uncertainty (not displayed in Figs.~\ref{protonbasis-u+d} 
and \ref{autsin2pppx-COM}) and is compatible with zero or opposite sign
within 1-$\sigma$. This TMD is difficult to measure in DY and SIDIS.
In the high luminosity SIDIS experiments at JLab 12 GeV and the 
future Electron Ion Collider it may be feasible to measure pretzelosity. 

The $A_{UU}^{\cos 2\phi}$ asymmetry in unpolarized DY originates 
from a convolution of the Boer-Mulders functions in nucleon and pion.
Historically it was connected to the ``violation'' of the Lam-Tung 
relation, see \cite{Boer:1999mm} and references therein.
A simultaneous measurement of $A_{UU}^{\cos 2\phi}$ and  
$A_{UT}^{\sin(2\phi-\phi_S)}$ which we have discussed above
allows one to test the sign change of the proton Boer-Mulders 
function in DY. $A_{UU}^{\cos 2\phi}$ was measured and found
positive in earlier CERN and Fermilab measurements
\cite{Guanziroli:1987rp,Conway:1989fs}.
Neglecting sea quark effects, the asymmetry is dominated by 
$A_{UU}^{\cos 2\phi}\propto h_{1, \pi^-}^{\perp (1)\bar u }(x_\pi) h_{1,p}^{\perp (1) u} (x_p)$. 
With the indication of the positive sign for the pion Boer-Mulders 
function from the COMPASS data \cite{Aghasyan:2017jop} on  
$A_{UT}^{\sin(2\phi-\phi_S)}$, we conclude a positive sign 
also for the proton $u$-quark Boer-Mulders function in DY, which 
is opposite to the sign seen in SIDIS analyses \cite{Barone:2010gk}
and hence in agreement with the prediction for the process dependence 
property of T-odd TMDs \cite{Collins:2002kn}.

Fig.~\ref{auucos2px-COM} shows our predictions for $A_{UU}^{\cos 2\phi}$ 
for COMPASS kinematics. At this point no data are available from 
COMPASS, but an analysis is planned~\cite{Gautheron:2010wva}
and our predictions in Fig.~\ref{auucos2px-COM} may be tested in near future. 
It is worth recalling that our approach provides a good description 
of the NA10 CERN \cite{Guanziroli:1987rp} and E615 Fermilab
\cite{Conway:1989fs} data. The test of our predictions 
in Fig.~\ref{auucos2px-COM} will help to investigate the compatibility 
of the NA10, E615 and COMPASS experiments.  
Interestingly, fixed-order collinear factorized perturbative 
QCD calculations, which strictly speaking require $q_T$ to be 
the hard scale, can also qualitatively describe 
the NA10 and E615 data \cite{Lambertsen:2016wgj,Chang:2018pvk}. 
It will be interesting to confront those calculations with future 
COMPASS data and TMD studies.

Notice that in the analysis \cite{Barone:2010gk}
of the proton-proton and proton-deuteron data from the FNAL 
E866/NuSea experiment \cite{Zhu:2006gx,Zhu:2008sj} indications
were obtained that the proton quark and antiquark Boer-Mulders 
functions (in DY) have the same signs.
 With our observations 
based on COMPASS data we therefore infer a first hint that 
also the Boer-Mulders functions of $\bar{u}$ and $\bar{d}$ 
are positive in DY. Interestingly, not only valence 
Boer-Mulders distributions in nucleon and pion seem ``alike''
\cite{Burkardt:2007xm}, but also the nucleon sea quark
distributions seem to have all the same sign. 
This confirms an early estimate on the sign
of the anti-quark Boer Mulders function carried in the SPM  in Ref.~\cite{Gamberg:2005ip}.
This is in line with
predictions from the limit of a large number of colors $N_c$ in QCD
that $h_{1,p}^{\perp u} (x_p,\kTN) = h_{1,p}^{\perp d} (x_p,\kTN)$ and 
$h_{1,p}^{\perp\bar u} (x_p,\kTN) = h_{1,p}^{\perp\bar d} (x_p,\kTN)$
modulo $1/N_c$ corrections \cite{Pobylitsa:2003ty}. 
Future data will provide more stringent tests of these 
predictions. 

Finally,  it is worth pointing out that in principle one can extract the
$u$-quark transversity distribution entirely from the measurements 
of $A_{UU}^{\cos 2\phi}$ and $A_{UT}^{\sin(2\phi-\phi_S)}$ 
in $\pi^-$-proton DY at COMPASS \cite{Sissakian:2005yp}.
While typically data available from different processes are
processed in ``global analyses,'' whenever possible it is
also valuable to extract a function from one process alone.
This would for instance allow one to test the universality 
(same sign and $x$-shape in SIDIS and DY) of the $u$-quark 
transversity distribution which is otherwise taken for granted.

Fig.~\ref{aulsin2px-COM} displays our predictions for the
longitudinal single-spin asymmetry $A_{UL}^{\sin 2\phi}$ 
in the COMPASS kinematics which is due to the 
Kotzinian-Mulders TMD $h_{1L}^{\perp a}$ and the pion 
Boer-Mulders function. If we disregard sea quark effects, then 
$A_{UL}^{\sin 2\phi}\propto  -h_{1, \pi^-}^{\perp (1)\bar u }(x_\pi) h_{1L,p}^{\perp (1) u} (x_p) > 0$. Especially the SPM predicts
a sizable and positive asymmetry. Since no parametrization on
$h_{1L}^{\perp a}$ is currently available, the hybrid calculations 
make use of the WW-type approximation which is compatible with 
SIDIS data \cite{Bastami:2018xqd}.
This is the only leading-twist pion-proton asymmetry in DY
which requires a longitudinal proton polarization.
We are not aware of plans to run DY experiments with longitudinal 
proton polarization in the near future. Potentially $A_{UL}^{\sin2\phi}$ 
could be studied in DY with doubly polarized protons or deuterons in a 
future NICA experiment \cite{Savin:2015paa}.

\begin{table}[t]
\centering
	\begin{tabular}{|c|c|c|c|c|c|}
		\hline
		Fig.\ & structure function & TMDs & LFCQM & SPM & phenomenology \\
		\hline \hline
		\ref{aut1x-COM}-\ref{aulsin2px-COM}       & $F_{UU}^1$ & $f_{1,p}^{a}$, \ $f_{1,\pi}^a$ &  \cite{Pasquini:2008ax}, \cite{Pasquini:2014ppa} & \cite{Gamberg:2007wm} \cite{Gamberg:2009uk} & \cite{Martin:2009iq}, \cite{Sutton:1991ay} \\
		\hline
		\ref{aut1x-COM} & $F_{UT}^{\sin\phi_S}$ & $f_{1T,p}^{\perp a}$, \ $f_{1,\pi}^a$ & \cite{Pasquini:2010af}, \cite{Pasquini:2014ppa} & \cite{Gamberg:2007wm}, \cite{Gamberg:2009uk} & \cite{Cammarota:2020qcw}, \cite{Sutton:1991ay} \\
		\hline
		\ref{autsin2pmpx-COM} & $F_{UT}^{\sin(2\phi-\phi_S)}$ & $h_{1,p}^{a}$, \ $h_{1,\pi}^{\perp a}$ & \cite{Pasquini:2008ax}, \cite{Pasquini:2014ppa} & \cite{Gamberg:2007wm}, \cite{Gamberg:2009uk} & \cite{Cammarota:2020qcw}, \ --- \\
		\hline
		\ref{autsin2pppx-COM} & $F_{UT}^{\sin(2\phi+\phi_S)}$ & $h_{1T,p}^{\perp a}$, \ $h_{1,\pi}^{\perp a}$ & \cite{Pasquini:2008ax}, \cite{Pasquini:2014ppa} & \cite{Jakob:1997wg}, \cite{Gamberg:2009uk} & \cite{Lefky:2014eia}, \ ---  \\
		\hline
		\ref{auucos2px-COM} & $F_{UU}^{\cos 2\phi}$ & $h_{1,p}^{\perp a}$, \ $h_{1,\pi}^{\perp a}$ & \cite{Pasquini:2010af}, \cite{Pasquini:2014ppa} & \cite{Gamberg:2007wm}, \cite{Gamberg:2009uk}
		& \cite{Barone:2009hw}, \ --- \\
		\hline
		\ref{aulsin2px-COM} & $F_{UL}^{\sin 2\phi}$ & $h_{1L,p}^{\perp a}$, \ $h_{1,\pi}^{\perp a}$ & \cite{Pasquini:2008ax}, \cite{Pasquini:2014ppa} & \cite{Gamberg:2007wm}, \cite{Gamberg:2009uk} & \cite{Bastami:2018xqd}, \ --- \\
		\hline
	\end{tabular}
	\caption{Overview on non-perturbative input used to produce 
	the results in Figs.~\ref{aut1x-COM}--\ref{aulsin2px-COM} which
	was taken from the LFCQM, the SPM, and phenomenological fits
	(or WW-type approximation in the case of
	$h_{1L,p}^{\perp a}$).
	Notice that no phenomenological information is currently available
	on $h_{1,\pi}^{\perp a}$, cf.\ section~\ref{sec-2.3}. 
	\label{table-overview}.}
\end{table}

\begin{figure}[t]
\centering
  \begin{tabular}{cc}
		 \rotatebox[origin=c]{90}{{$A_{UU}^{\cos 2\phi}$}} &  
		 \hspace{-2.5mm}\raisebox{-.5\height}{\includegraphics[width=5.5cm]{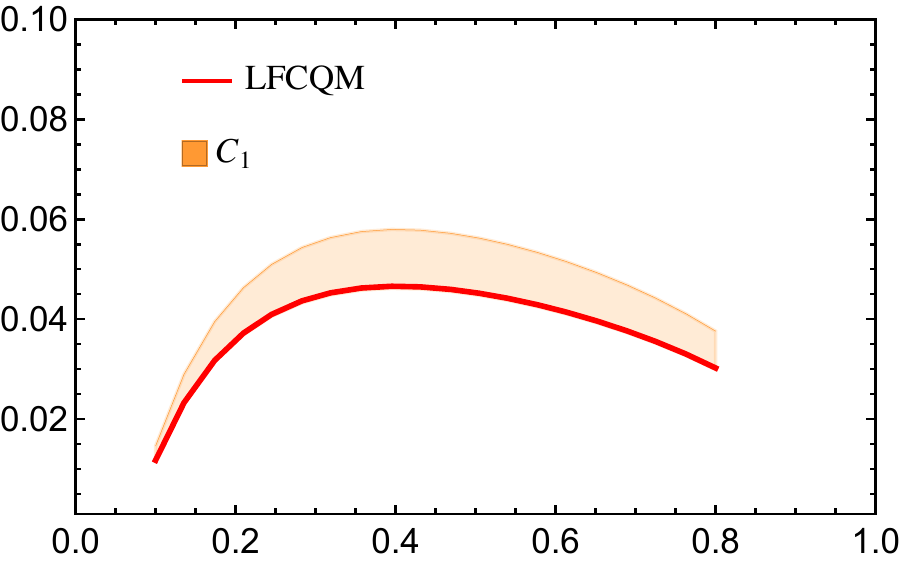}}  \\
		& $x_\pi$  \\	
  \end{tabular}
\vspace{-3.5mm}
\caption{\label{auucos2px-COM-study} 
 $A_{UU}^{\cos 2\phi}$ as a function of 
 $x_{\pi}$, the orange region corresponds to variation of $C_1$ in the interval $ [e^{-\gamma_E}, 4\, e^{-\gamma_E}]$ and illustrates the sensitivity of our results to scale variations.}
\end{figure}

We also study the  theoretical uncertainty  due to the variation of $C_1$ and $C_2$ in Eqs.~(\ref{eq:hard}), (\ref{e:FF_ansatz}), and (\ref{e:CS_kernel0})  at NLL accuracy. Such studies are of importance in order to establish the control over the perturbative expansion, see e.g. \cite{Scimemi:2018xaf}. We will use $A_{UU}^{\cos 2\phi}$ asymmetry as an example. Scale dependence on $C_2$ cancels exactly at this order between the numerator and the denominator of the asymmetry. In Fig.~\ref{auucos2px-COM-study} we show the corresponding theoretical uncertainty due to variation of $C_1\in [e^{-\gamma_E}, 4\,e^{-\gamma_E}]$ for the LFCQM, notice that for $C_1= e^{-\gamma_E}$ the asymmetry becomes larger than the red curve calculated with  $C_1= 2 e^{-\gamma_E}$,
 while for $C_1= 4 e^{-\gamma_E}$ the value of asymmetry decreases very slightly. One can see that the theoretical uncertainty due to the scale choice of $C_1$ is not negligible and warrants  the inclusion of higher order corrections in the calculations. This uncertainty is smaller than the spread of the model predictions shown in Fig.~\ref{auucos2px-COM} and therefore we expect that the future data will be able to distinguish among various models.

Before ending this section it is important to remark that the COMPASS
experiment has covered the range $0.4\,{\rm GeV} < q_T < 5\,{\rm GeV}$. 
At the upper limit the condition $q_T \ll  Q$ for
the applicability of the TMD factorization is not satisfied which 
constitutes an  uncertainty in our calculations.
However, in the experiment (and in our calculations) it is 
$\langle q_T\rangle = 1.2 \,{\rm GeV}$ which is much smaller 
than $\langle Q \rangle = 5.3\,{\rm GeV}$ and we verified that 
the region of large $q_T$ (namely, $3\,{\rm GeV} < q_T < 5\,{\rm GeV}$) 
in our calculations has a negligible impact on
the $q_T$-averaged (integrated) asymmetries in the experiment.

\section{Conclusions}
\label{sec-4}

In this work we studied the DY process with negative pions 
and polarized protons with focus on the kinematics of the 
COMPASS experiment. As no phenomenological extractions are 
available for the Boer-Mulders TMD
function of the pion, we explored two popular and widely used
hadronic models, the LFCQM and the SPM, together with available
phenomenological information on the other TMDs. For the LFCQM and SPM the we implement TMD evolution at NLL accuracy from fixed scale according to the solution to the CSS equations in Ref.~\cite{Collins:2014jpa} and outlined in section~\ref{sec-2.2}. This approach moves beyond the approximate TMD evolution based on the Gaussian Ansatz for transverse parton momenta with
energy dependent Gaussian widths.

We presented a complete description of polarized DY at leading 
twist using TMD evolution at NLL accuracy. 
The required TMDs include on the nucleon side
$f_{1,p}^{a}$, $f_{1T,p}^{\perp a}$,
$h_{1,p}^{a}$, $h_{1,p}^{\perp a}$, 
$h_{1T,p}^{\perp a}$, $h_{1L,p}^{\perp a}$;
and on the pion side
$f_{1,\pi}^a$, $h_{1,\pi}^{\perp a}$.
For that we compiled results from several prior LFCQM and SPM
calculations, which to the best of our knowledge have not been 
presented in this completeness before
\cite{Jakob:1997wg,Pasquini:2010af,Pasquini:2008ax,Pasquini:2014ppa,Gamberg:2007wm,Gamberg:2009uk}. 
Based on concise comparisons of model results with 
available phenomenological information 
\cite{Martin:2009iq,Sutton:1991ay,Anselmino:2011gs,Anselmino:2013vqa,Barone:2009hw,Lefky:2014eia,Cammarota:2020qcw,Bastami:2018xqd},
we estimate an accuracy of the model results of
20-40$\,\%$ for the majority of (though not all) TMDs. Similar  
``model accuracies'' were found  in prior phenomenological 
applications of CQMs \cite{Boffi:2009sh,Pasquini:2011tk,Pasquini:2014ppa}.

Driven by the motivation to make maximal use of currently 
available phenomenological information
\cite{Martin:2009iq,Sutton:1991ay,Anselmino:2011gs,Anselmino:2013vqa,Barone:2009hw,Lefky:2014eia,Cammarota:2020qcw,Bastami:2018xqd}, 
we also carried out ``hybrid'' calculations with a minimal model 
dependence --- namely only due to the pion Boer-Mulders function 
for which no extraction is currently available. 
In this way we provided up to four predictions 
for each DY observable, with different levels of model dependence. 
The critical comparison of the various results 
(pure-model and hybrid calculations in respectively LFCQM and SPM) 
allows us to differentiate robust predictions from more strongly
model-dependent results.

Our study had two main goals, namely to present theoretical 
calculations which help to interpret the first data from the
pion-induced DY with polarized protons measured by COMPASS, 
as well as to provide quantitative tests of 
the application of 
CQMs to the description of pion and nucleon structure.

In regard to the interpretation of the first data from the
pion-induced DY with polarized protons, we observe a robust
picture. The pure-model and hybrid calculations from the
LFCQM and SPM are in remarkable agreement with each other
at the present stage. The theoretical spread of our
results is smaller than the present uncertainties of the
available data. Among the most interesting observations are
the encouraging indications for the change of sign of the
T-odd TMDs in DY vs SIDIS, both in the case of the proton
Sivers and proton Boer-Mulders function. These are model
independent results. Another model-independent result is
the observation that the data favor a positive (in DY)
Boer-Mulders $\bar u$-distribution in $\pi^-$.
We also report the first indication that all proton
Boer-Mulders functions for $u$, $d$, $\bar u$, $\bar d$
flavors are positive (in DY). At the present, these 
observations are admittedly vague due to the  
low precision of the 
current data. More precise
future data from COMPASS and other facilities will allow us 
to solidify the picture.

In regard to the quantitative tests of the application 
of CQMs, it is important to stress that the DY process
with $\pi^-$ and proton in the COMPASS kinematics is an ideal 
process for these purposes. In the COMPASS kinematics sea quarks 
do not play an important role \cite{Collins:2005rq}. Due to the 
$u$-quark dominance in the proton the process is strongly dominated 
by annihilations of $\bar u$ from $\pi^-$ and $u$ 
from proton in the valence $x$-region
where CQMs can be expected to catch the main features in the hadronic structure of the 
pion and nucleon. 

CQMs are important qualitative tools for QCD calculations. 
Within their model accuracy 
and within their range of applicability 
in the valence $x$-region,
we observe that CQMs  yield useful results and provide 
helpful guidelines for the interpretation of data. Future data 
will provide more stringent tests of the CQMs, and allow 
for extraction of  hadron structure 
by global QCD analyses. 
We also provided several predictions that await experimental
confirmation.

\section*{Acknowledgments}
The authors wish to thank A.~V.~Efremov and A. Kotzinian for valuable discussions 
which motivated this study and J.~Collins, T.~Rogers, and Z.~Kang for discussions on implementation of TMD evolution.
This work was supported by the National Science Foundation under the
Contracts No.\ 
PHY-1812423 (S.B.\ and P.S.) and No.~PHY-2012002 (A.P.), 
and in part by the US Department of Energy under contracts, No.~DE-FG02-07ER41460 (L.G.) 
and No.~DE-AC05-06OR23177 (A.P.) under which JSA, LLC operates JLab, the framework of the TMD Topical Collaboration (L.G. and A.P.), and by the European Union's Horizon 2020 program under grant agreement No.~824093(STRONG2020) (B.P.).

\bibliography{\BibPath/biblio_DY}

\end{document}